\documentclass[a4paper, 11pt]{article}\usepackage[]{graphicx}\usepackage[dvipsnames,table]{xcolor}
\makeatletter
\def\maxwidth{ %
  \ifdim\Gin@nat@width>\linewidth
    \linewidth
  \else
    \Gin@nat@width
  \fi
}
\makeatother

\definecolor{fgcolor}{rgb}{0.345, 0.345, 0.345}
\newcommand{\hlnum}[1]{\textcolor[rgb]{0.686,0.059,0.569}{#1}}%
\newcommand{\hlsng}[1]{\textcolor[rgb]{0.192,0.494,0.8}{#1}}%
\newcommand{\hlcom}[1]{\textcolor[rgb]{0.678,0.584,0.686}{\textit{#1}}}%
\newcommand{\hlopt}[1]{\textcolor[rgb]{0,0,0}{#1}}%
\newcommand{\hldef}[1]{\textcolor[rgb]{0.345,0.345,0.345}{#1}}%
\newcommand{\hlkwb}[1]{\textcolor[rgb]{0.69,0.353,0.396}{#1}}%
\newcommand{\hlkwc}[1]{\textcolor[rgb]{0.333,0.667,0.333}{#1}}%
\newcommand{\hlkwd}[1]{\textcolor[rgb]{0.737,0.353,0.396}{\textbf{#1}}}%

\usepackage{framed}
\makeatletter
\newenvironment{kframe}{%
 \def\at@end@of@kframe{}%
 \ifinner\ifhmode%
  \def\at@end@of@kframe{\end{minipage}}%
  \begin{minipage}{\columnwidth}%
 \fi\fi%
 \def\FrameCommand##1{\hskip\@totalleftmargin \hskip-\fboxsep
 \colorbox{shadecolor}{##1}\hskip-\fboxsep
     \hskip-\linewidth \hskip-\@totalleftmargin \hskip\columnwidth}%
 \MakeFramed {\advance\hsize-\width
   \@totalleftmargin\z@ \linewidth\hsize
   \@setminipage}}%
 {\par\unskip\endMakeFramed%
 \at@end@of@kframe}
\makeatother

\definecolor{shadecolor}{rgb}{.97, .97, .97}
\definecolor{messagecolor}{rgb}{0, 0, 0}
\definecolor{warningcolor}{rgb}{1, 0, 1}
\definecolor{errorcolor}{rgb}{1, 0, 0}
\newenvironment{knitrout}{}{} 

\usepackage{alltt}
\usepackage{amsmath, amssymb}
\usepackage{doi} 
\usepackage[round]{natbib} 
\usepackage{multirow}
\usepackage{booktabs} 
\usepackage[title]{appendix} 
\usepackage{nameref} 
\usepackage[dvipsnames,table]{xcolor}
\usepackage[onehalfspacing]{setspace}
\usepackage[labelfont=bf,font=small]{caption} 
\usepackage{helvet}
\usepackage{mathpazo}
\usepackage{sectsty} 
\allsectionsfont{\sffamily} 
\usepackage[labelfont={bf,sf},font=small]{caption} 
\usepackage{orcidlink} 
\definecolor{lightgray}{gray}{0.9}
\usepackage{pifont} 
\newcommand{\cmark}{\ding{51}}%
\newcommand{\xmark}{\ding{55}}%

\newif\ifanonymize 
\newcommand{\anonymize}[1]{%
  \ifanonymize
    \phantom{#1}%
  \else
    #1%
  \fi
}
\anonymizefalse 

\usepackage{geometry}
\newcommand\mail{samuel.pawel@uzh.ch}
\title{\vspace{-4em}
  \textbf{\textsf{Closed-Form Power and Sample Size Calculations for Bayes Factors}}
}
\author{
  \anonymize{\textbf{Samuel Pawel} \orcidlink{0000-0003-2779-320X}} \and
  \anonymize{\textbf{Leonhard Held} \orcidlink{0000-0002-8686-5325}}
}
\date{
  \anonymize{Epidemiology, Biostatistics and Prevention Institute (EBPI)} \\
  \anonymize{Center for Reproducible Science (CRS)} \\
  \anonymize{University of Zurich} \\
  \anonymize{E-mail: \href{mailto:\mail}{\mail}} \\[2ex]
  \anonymize{{\color{blue} Preprint version November 13, 2024}}
}

\usepackage{hyperref}  
\hypersetup{
  bookmarksopen=true, 
  breaklinks=true,
  pdftitle={Closed-Form Power and Sample Size Calculations for Bayes Factors},
  pdfsubject={},
  pdfkeywords={},
  colorlinks=true,
  linkcolor=black,
  anchorcolor=black,
  citecolor=blue,
  urlcolor=blue,
}

\usepackage{fancyhdr}
\IfFileExists{upquote.sty}{\usepackage{upquote}}{}
\begin{document}
\maketitle

\begin{abstract}
  \noindent Determining an appropriate sample size is a critical element of study design, and the method used to determine it should be consistent with the planned analysis. When the planned analysis involves Bayes factor hypothesis testing, the sample size is usually desired to ensure a sufficiently high probability of obtaining a Bayes factor indicating compelling evidence for a hypothesis, given that the hypothesis is true. In practice, Bayes factor sample size determination is typically performed using computationally intensive Monte Carlo simulation. Here, we summarize alternative approaches that enable sample size determination without simulation. We show how, under approximate normality assumptions, sample sizes can be determined numerically, and provide the R package \texttt{bfpwr} for this purpose. Additionally, we identify conditions under which sample sizes can even be determined in closed-form, resulting in novel, easy-to-use formulas that also help foster intuition, enable asymptotic analysis, and can also be used for hybrid Bayesian/likelihoodist design. Furthermore, we show how power and sample size can be computed without simulation for more complex analysis priors, such as Jeffreys-Zellner-Siow priors or non-local normal moment priors. Case studies from medicine and psychology illustrate how researchers can use our methods to design informative yet cost-efficient studies.
  \\
  \textit{Keywords:} Bayesian hypothesis testing, design prior, evidence,
  likelihood ratio, study design
\end{abstract}

\section{Introduction}
A key aspect of study design is determining an appropriate sample size. Choosing
a sample size that is too small may lead to inconclusive study results, while
choosing a sample size that is too large may be unethical (e.g., for animal
studies) or waste samples which could be of better use in other studies. Whether
or not a certain sample size can ensure sufficiently conclusive results depends
on the planned analysis. Therefore, the sample size calculation should be
aligned with the goals of the analysis \citep{Anderson2022}, or in other words:
`\emph{As ye shall analyse is as ye shall design}' \citep[p.~179]{Julious2023}.

A widely used formula for the sample size per group (with two groups of equal
size) for continuous outcome data, based on a frequentist hypothesis test of a
mean difference, is given by
\begin{align}
  \label{eq:freqsamplesize}
  n = \frac{2 \sigma^{2} (z_{1 - \alpha/2} + z_{1 - \beta})^2}{(\mu - \theta_0)^2},
\end{align}
where $z_{q}$ is the $q\times 100\%$ quantile of the standard normal
distribution, $\alpha$ is the level of the test, $1 - \beta$ is the desired
power, $\mu$ is the assumed mean difference, $\theta_0$ is the mean difference
under the null hypothesis, and $\sigma^2$ is the variance of one observation
\citep[p.~34]{Matthews2006}. There exist various refinements
of~\eqref{eq:freqsamplesize}, such as, adaptations to unequal randomization,
special study designs (e.g., cross-over studies), or other data types (e.g.,
binary data), see, for example, \citet{Kieser2020} or \citet{Julious2023}. For
many analysis methods, however, no closed-form formula exist and iterative or
simulation methods have to be used. Nevertheless, while typically only being an
approximation, the formula~\eqref{eq:freqsamplesize} enables `quick-and-dirty'
calculations that are often accurate enough for practical purposes. It also
helps fostering intuition and is therefore useful, for example, in teaching of
statistics. Finally, the formula is helpful for theoretical derivations such as
`how much do we need to increase the sample size if we reduce $\alpha=0.05$ down
to $\alpha=0.005$?' \citep{Benjamin2017}.

An alternative to frequentist hypothesis testing is Bayesian hypothesis testing.
There are different flavors of Bayesian hypothesis testing, one of the most
popular being the approach centered around the \emph{Bayes factor}, which is the
data-based updating factor of the prior to posterior odds of two competing
hypotheses. Bayes factor approaches were pioneered by \citet{Jeffreys1939} and
are now in use in various scientific domains such as medicine
\citep{Goodman1999}, psychology \citep{Morey2016, Heck2023}, or physics
\citep{Trotta2008}. Bayes factor tests are conceptually different from
frequentist tests in several ways. For example, they can quantify evidence in
favor of a null hypothesis or they can incorporate external information via a
prior distribution. For an overview of Bayes factors see e.g., \citet{Kass1995,
  Held2018}.

Also if Bayes factors are used in the analysis, the design of the study should
match the analysis. Fortunately, there is methodology for design based on Bayes
factors \citep{Weiss1997, Gelfand2002, DeSantis2004, DeSantis2007,
  Schoenbrodt2017, Schnbrodt2017b, Pawel2023e, Stefan2022}. However, to our
knowledge, there are no simple formulas such as~\eqref{eq:freqsamplesize} for
sample size determination based on Bayes factor analyses. In practice, sample
size determination is often performed by Monte Carlo simulation
\citep{Gelfand2002, Schoenbrodt2017, Stefan2022}, but this can be inaccurate,
time-consuming, and less intuitive than a formula.

The goal of this paper is therefore to investigate whether, under approximate
normality assumptions similar to those underlying the
formula~\eqref{eq:freqsamplesize}, a sample size formula can be derived for a
planned Bayes factor analysis. As we will show, the answer is affirmative under
certain assumptions about the analysis prior (the prior distribution for the
parameter used in the analysis) and the design prior (the prior distribution for
the parameter used in the design). A distinction must be made between point
priors and normal priors, both in the design and in the analysis. Point analysis
priors lead to Bayes factors reducing to likelihood ratios, the analysis thereby
corresponding with a frequentist `likelihoodist' analysis \citep{Edwards1971,
  Royall1997, Blume2002, Strug2018, Cahusac2020}, while point design priors lead
to `conditional power', which corresponds to traditional frequentist power. In
contrast, normal analysis and design priors can account for parameter
uncertainty, producing Bayes factors that differ from likelihood ratios and
`predictive power' that differs from frequentist power \citep[see e.g.,][for
conditional/predictive power related to posterior tail probability
analyses]{OHagan2005, Micheloud2020, Grieve2022}.

Based on the point/normal prior distinction, we find that closed-form sample
sizes are available for Bayes factors with point analysis priors (i.e.,
likelihood ratios) along with point or normal design priors, and for Bayes
factors with local normal analysis and design priors (i.e., normal priors
centered on the null value). Table~\ref{tab:summary} provides an overview of our
results. In addition to our novel formulas, we summarize sample size
determination for Bayes factors with normal priors based on numerical
root-finding \citep[which has been done before, e.g., in][]{Weiss1997,
  Pawel2023e}, and show how it can be extended to more advanced prior
distributions, such as normal moment priors and Jeffreys-Zellner-Siow priors.
While root-finding approaches also do not produce closed-form sample sizes,
computations are deterministic and usually faster than with simulation
approaches. To facilitate reuse of our results, all methods are made available
through our R package \texttt{bfpwr}.

\begingroup
\renewcommand{\arraystretch}{1.3} 
\begin{table}[!htb]
  \centering
  \caption{Availability of closed-form sample size formulas for Bayes factor
    hypothesis test of $H_{0} \colon \theta = \theta_{0}$ against
    $H_{1} \colon \theta \neq \theta_{0}$ with either a point prior or a normal
    analysis prior assigned to $\theta$ under $H_{1}$, and data in the form of a
    normally distributed parameter estimate
    $\hat{\theta} \mid \theta \sim \mathrm{N}(\theta, \sigma_{\scriptscriptstyle \hat{\theta}}^{2}/n)$.
    In all cases, the power can be computed in closed-form.}
  \label{tab:summary}
  {\small
  \begin{tabular}{l c c }
    \toprule
    & \multicolumn{2}{c}{Analysis prior} \\
    \cmidrule{2-3}
    \multicolumn{1}{c}{Design prior} & Point prior (likelihood ratio) & Normal prior (Bayes factor) \\
    \midrule
    Point prior (conditional power) 
 &\cellcolor{green!10}  \cmark ~ Equation~\eqref{eq:nLR2} & \cellcolor{red!10} \xmark ~ Unavailable \\
    Normal prior (predictive power) 
    & \cellcolor{green!10} \cmark ~ Equation~\eqref{eq:nLR} & \cellcolor{yellow!10} \cmark ~ Equation~\eqref{eq:nBF} for local normal priors \\
    \bottomrule
  \end{tabular}
  }
\end{table}
\endgroup

This paper is organized as follows: We begin by defining the type of Bayes
factor underlying our sample size calculations (Section~\ref{sec:BF}), followed
by deriving its distribution when new data are generated under various design
priors (Section~\ref{sec:distBF}). Combining these results, we derive several
formulas for the sample size under different constellations of design and
analysis priors (Section~\ref{sec:ssd}). Examples from medicine and psychology
then illustrate how our formulas can be used by researchers in practice
(Section~\ref{sec:application}). Section~\ref{sec:extensions} illustrates
possible extensions of our framework to other popular types of analysis priors,
such as informed $t$ priors \citep{Gronau2020} and non-local normal moment priors
\citep{Johnson2010}. The paper ends with a closing discussion of our results and
final remarks on limitations and extensions (Section~\ref{sec:discussion}). Our
R package \texttt{bfpwr} that implements the developed methods is illustrated in
Appendix~\ref{app:pkg}.

\section{Bayes factor analysis}
\label{sec:BF}
To derive a sample size formula, we must first clarify how the future data will
be analyzed. Denote by $\hat{\theta}$ the estimate of an unknown parameter
$\theta$ that will result from the statistical analysis of a future data set
with effective sample size $n$. Suppose that the estimate's standard error is of
the form $\sigma_{\scriptscriptstyle \hat{\theta}}/\sqrt{n}$ where
$\sigma_{\scriptscriptstyle \hat{\theta}}^2$ is the variance of one effective
observation. For example, if the data are normally distributed with known
variance $\sigma^2$ and the parameter of interest is their mean $\theta$
estimated with the sample mean $\hat{\theta}$ (the `one-sample' case), we have
that $\mathrm{Var}(\hat{\theta}) = \sigma^2/n$ so the unit variance is
$\sigma_{\scriptscriptstyle \hat{\theta}}^2 = \sigma^2$ and the effective sample
size $n$ is the number of observations. If there were another group of $n$
normally distributed observations with a potentially different mean but the same
known variance $\sigma^2$, and we were interested in the mean difference
$\theta$ estimated with the empirical mean difference $\hat{\theta}$ (the
`two-sample' case), we would have that $\mathrm{Var}(\hat{\theta}) =
(2\sigma^2)/n$ so the unit variance would be $\sigma_{\scriptscriptstyle
  \hat{\theta}}^2 = 2\sigma^2$ and the effective sample size $n$ would be the number
of observations per group. For many commonly used estimators, it is
reasonable to assume that the estimate (but not necessarily the underlying data)
is approximately normally distributed around $\theta$ with variance equal to the
squared standard error, i.e., \mbox{$\hat{\theta} \mid \theta \sim
  \mathrm{N}(\theta, \sigma^2_{\scriptscriptstyle \hat{\theta}}/n)$}.
Table~\ref{tab:outcomes} shows common types of parameter estimates and the
resulting interpretation of the effective sample size $n$ and the unit variance
$\sigma^2_{\scriptscriptstyle \hat{\theta}}$. It is important to note that these
are only approximations and they may be inadequate in certain situations, such
as small sample sizes or when the assumed variance $\sigma^2$ is strongly
misspecified \citep{Spiegelhalter2004}. The frequentist sample size
formula~\eqref{eq:freqsamplesize} can also be cast in this framework; it assumes
continuous outcome data and a mean difference parameter (second row in
Table~\ref{tab:outcomes}). As such, the formula could be generalized to other
settings by replacing $2\sigma^{2}$ in the numerator with other unit variances
from Table~\ref{tab:outcomes}, which would in turn change the interpretation of
$n$ and $\hat{\theta}$.

\begingroup
\renewcommand{\arraystretch}{1.3} 
\begin{table}[!htb]
  \centering
  \caption{Different types of parameter estimates $\hat{\theta}$ with
    approximate variance $\mathrm{Var}(\hat{\theta}) =
    \sigma^2_{\scriptscriptstyle \hat{\theta}}/n$ and corresponding
    interpretation of sample size $n$ and unit variance
    $\sigma^2_{\scriptscriptstyle \hat{\theta}}$ (adapted from Chapter 2.4 in
    \citealp{Spiegelhalter2004} and Chapter 1 in \citealp{Grieve2022}). The
    variance of one continuous outcome observation is denoted by $\sigma^{2}$
    and assumed to be known. Parameter estimates based on two groups assume an
    equal number of observations per group.}
  \label{tab:outcomes}
  \rowcolors{1}{}{lightgray}
  \begin{tabular}{l l l c}
    \toprule
    \textbf{Outcome} & \textbf{Parameter estimate} $\hat{\theta}$ & \textbf{Interpretation of} $n$ & \textbf{Unit variance} $\sigma^2_{\scriptscriptstyle \hat{\theta}}$ \\
    \midrule
    Continuous & Mean & Sample size & $\sigma^{2}$ \\
    Continuous & Mean difference & Sample size per group & $2\sigma^{2}$ \\
    Continuous & Standardized mean difference & Sample size per group & 2 \\
    Continuous & $z$-transformed correlation & Sample size minus 3 & 1 \\
    Binary & Arcsine square root difference & Sample size per group & 1/2 \\
    Binary & Log odds ratio & Total number of events & 4 \\
    Survival & Log hazard ratio & Total number of events & 4 \\
    Count & Log rate ratio & Total count & 4 \\

    \bottomrule
    \end{tabular}
\end{table}
\endgroup

Assume now a point null hypothesis that postulates that $\theta$ equals a
certain null value \mbox{$H_0 \colon \theta = \theta_0$} and an alternative
hypothesis that postulates that $\theta$ does not equal the null value
$H_1 \colon \theta \neq \theta_0$, with prior
\mbox{$\theta \mid H_1 \sim \mathrm{N}(\mu, \tau^2)$} assigned to $\theta$ under
$H_1$. The mean of the prior $\mu$ determines the most plausible parameter value
under the alternative while the standard deviation $\tau$ determines its
uncertainty. A point alternative at $\mu$ may be obtained by letting the
standard deviation of the prior go to zero. Whenever we write $\tau = 0$, we
informally refer to a point prior at $\mu$, since for all calculations in this
paper this notation leads to the same results as a more formal treatment of
point priors. The Bayes factor is then given by the updating factor of the prior
to posterior odds of $H_0$ versus $H_1$, i.e.,
\begin{align}
\label{eq:BF01}
    \text{BF}_{01} 
    = \frac{\Pr(H_0 \mid \hat{\theta})}{\Pr(H_1 \mid \hat{\theta})} \bigg/ \, \frac{\Pr(H_0)}{\Pr(H_1)}
    = \sqrt{1 + \frac{n \tau^2}{\sigma^2_{\scriptscriptstyle \hat{\theta}}}} \exp\left[-\frac{1}{2} \left\{\frac{(\hat{\theta} - \theta_0)^2}{\sigma^2_{\scriptscriptstyle \hat{\theta}}/n} - \frac{(\hat{\theta} - \mu)^2}{\tau^2 + \sigma^2_{\scriptscriptstyle \hat{\theta}}/n}\right\}\right].
\end{align}
A Bayes factor less than one ($\text{BF}_{01} < 1$) indicates evidence for the
alternative hypothesis $H_{1}$, while a Bayes factor greater than one
($\text{BF}_{01} > 1$) indicates evidence for the null hypothesis $H_{0}$. The
larger the deviation of the Bayes factor from one, the stronger the evidence.
Conventional thresholds for substantial and strong evidence for the alternative
(null) hypothesis are $1/3$ and $1/10$ ($3$ and $10$), respectively. A Bayes
factor $1/3 < \text{BF}_{01} < 3$ is typically interpreted as absence of
evidence for either hypothesis, calling for more data to discern the more
appropriate hypothesis \citep{Jeffreys1939, Held2018}.

The Bayes factor~\eqref{eq:BF01} is implemented in our package in the function
\texttt{bf01}. It has already appeared in the literature in one form or another
\citep[e.g., in][]{Weiss1997, DeSantis2004, Spiegelhalter2004, Dienes2014,
  Bartos2023b}, with perhaps the first proposal of a Bayes factor based on an
approximately normally distributed parameter estimate and its standard error
dating back to \citet{Jeffreys1936}, see also \citet{Wagenmakers2022} for some
historical notes on Jeffreys' approach. The Bayes factor~\eqref{eq:BF01} may be
thought of as a `Bayesian \mbox{$z$-test}', that is, a test of a normal mean
based on an asymptotically normal statistic assuming that the variance of the
statistic is known. Of course, the latter assumption is not true in most
applications, but it makes the test widely applicable and is, for practical
purposes, often close enough to Bayes factors based on the exact distribution of
the data, which may or may not be available. Finally, the Bayes
factor~\eqref{eq:BF01} can also be used with parameter estimates where a
standard error is available but not of the form
$\sigma_{\scriptscriptstyle \hat{\theta}}/\sqrt{n}$, e.g., a parameter estimate
from a generalized linear model where the estimate is adjusted for covariates
and the standard error is obtained numerically. In this case,
$\sigma_{\scriptscriptstyle \hat{\theta}}/\sqrt{n}$ in~\eqref{eq:BF01} can be
replaced by the observed standard error. However, as we will show now, assuming
such a particular dependence on the sample size $n$ allows us to perform
closed-form power and sample size calculations under certain additional
assumptions.

\section{Distribution and power function of the Bayes factor}
\label{sec:distBF}
Suppose now that we are interested in finding compelling evidence -- either in
favor of the alternative $H_{1}$ over the null hypothesis $H_{0}$ with a Bayes
factor~\eqref{eq:BF01} smaller than some threshold $k < 1$ (e.g., $k = 1/3$ or
$k = 1/10$) or in favor of $H_{0}$ over $H_{1}$ with a Bayes factor greater than
$k > 1$ (e.g., $k = 3$ or $k = 10$).
To determine a sample size that ensures compelling evidence with a desired
probability we need to know the distribution of the Bayes factor~\eqref{eq:BF01}
for a given sample size.

Assume a so-called `design prior' for the parameter $\theta$ that is used in the
design of the study \citep{OHagan2001b, OHagan2005}. This prior should represent
the state of knowledge and uncertainty about $\theta$ at the design stage and
does not necessarily have to correspond to the `analysis prior' \mbox{$\theta
  \mid H_{1} \sim \mathrm{N}(\mu, \tau^{2})$} used for the Bayes
factor~\eqref{eq:BF01}. In fact, the analysis prior is often set to a certain
`default' or `objective' prior that is conventionally used in the field. Here,
we will focus on normal design priors $\theta \sim \mathrm{N}(\mu_{d},
\tau^{2}_{d})$, as they are flexible enough to specify varying degrees of
uncertainty about the parameter, and at the same time mathematically convenient
for obtaining closed-form solutions for power and, in some cases, sample size.
Point priors, and as such classical sample size determination with an `assumed
parameter', then represent a special case of the normal prior where the standard
deviation becomes infinitesimally small ($\tau_{d} = 0$). In addition,
specifying a point design prior at the null value ($\mu_{d} = \theta_{0}$ and
$\tau_{d} = 0$) allows us to compute the probability of finding compelling
evidence for the true null hypothesis $H_{0}$, as well as the probability of
finding misleading evidence for the alternative when the null hypothesis is true
(the `type I error rate'). Presenting the latter probability can be useful for
demonstrating that the chosen design is appropriately `calibrated'
\citep{Dawid1982, Rubin1984, Little2006, Grieve2016}.

Such a normal design prior induces a predictive distribution $\hat{\theta} \mid
n, \mu_{d} , \tau_{d} \sim \mathrm{N}(\mu_{d}, \tau^{2}_{d} +
\sigma^{2}_{\scriptscriptstyle \hat{\theta}}/n)$ for the future parameter
estimate $\hat{\theta}$, which is again a normal distribution centered around
the design prior mean $\mu_{d}$ but with a variance given by the sum of the
squared standard error $\sigma^{2}_{\scriptscriptstyle \hat{\theta}}/n$ and the
design prior variance $\tau_{d}^{2}$. Under this distribution, the distribution
of the Bayes factor with normal analysis prior~\eqref{eq:BF01} can be derived in
closed-form \citep{Weiss1997, DeSantis2004}. We now rederive and extend this
result in our setting and notation. The two cases of the Bayes factor with $\tau
= 0$ (point analysis prior under the alternative) and $\tau > 0$ (normal
analysis prior under the alternative) need to be distinguished, as the resulting
Bayes factor distributions take a different form, and only the latter has been
considered previously in the Bayes factor literature.

For the Bayes factor with point analysis prior ($\tau = 0$), the cumulative
distribution or `power function' is
\begin{align}
\label{eq:prLR}
  \Pr(\text{BF}_{01} \leq k \mid n, \mu_{d} , \tau_{d}, \tau = 0)
  = \begin{cases}
      1 - \Phi(Z) & \text{if}~ \mu - \theta_0 > 0 \\
      \Phi(Z) & \text{if}~ \mu - \theta_0 < 0
    \end{cases}
\end{align}
with $\Phi(\cdot)$ the standard normal cumulative distribution function and
\begin{align}
  \label{eq:Z}
    Z = \frac{1}{\sqrt{\tau^{2}_{d} + \sigma^{2}_{\scriptscriptstyle \hat{\theta}}/n}}\left\{
    \frac{\sigma^2_{\scriptscriptstyle \hat{\theta}}\log k}{n(\theta_0 - \mu)} + \frac{\theta_0 + \mu}{2}
    - \mu_{d}\right\},
\end{align}
see Appendix~\ref{app:distributions} for details. We may also want to compute
the probability that the Bayes factor is greater than $k$ (e.g., to determine
the probability of compelling evidence for a true null hypothesis), which is
simply one minus the probability~\eqref{eq:prLR}.

In standard frequentist sample size determination, the power can typically be
increased arbitrarily close to one by increasing the sample size.
However, with the Bayes factor based on a point analysis prior, depending on the
assumed design prior, one may not be able to approach a power of one with the
power function~\eqref{eq:prLR} by increasing the sample size $n$. That is, the
limiting power value is given by
\begin{align}
  \label{eq:powerlimLR}
  \lim_{n \to \infty} \Pr(\mathrm{BF}_{01} \leq k \mid n, \mu_{d} , \tau_{d}, \tau = 0)
    = \begin{cases}
      1 - \Phi(Z_{\mathrm{lim}}) & \text{if}~ \mu - \theta_0 > 0 \\
      \Phi(Z_{\mathrm{lim}}) & \text{if}~ \mu - \theta_0 < 0
    \end{cases}
\end{align}
with $Z_{\mathrm{lim}} = (\theta_{0} + \mu - 2 \mu_{d})/(2 \tau_{d})$, the limit
of~\eqref{eq:Z} for increasing $n$. When the design prior is also a point prior
($\tau_{d} = 0$), $Z_{\mathrm{lim}}$~diverges and the limiting
power~\eqref{eq:powerlimLR} approaches one or zero, depending on whether the
location of the design prior $\mu_{d}$ is closer to the alternative $\mu$ or to
the null $\theta_{0}$. In case it is just in between the two ($\mu_{d} =
(\theta_{0} + \mu)/2$), the limiting power approaches a half. On the other hand,
for a normal design prior ($\tau_{d} > 0$), the limiting power is bounded by a
value in between (and not including) zero and one given
by~\eqref{eq:powerlimLR}. The intuition behind these results is that for a
normal design prior, there is always parameter uncertainty, even if the sample
size becomes arbitrarily large, while for point design priors, the parameter
uncertainty can be arbitrarily reduced by increasing the sample size. This
parallels similar results on bounds for hybrid Bayesian/frequentist power
\citep{Spiegelhalter2004, Micheloud2020, Grieve2022}.

\begin{figure}[!tb]
\begin{knitrout}
\definecolor{shadecolor}{rgb}{0.969, 0.969, 0.969}\color{fgcolor}
\includegraphics[width=\maxwidth]{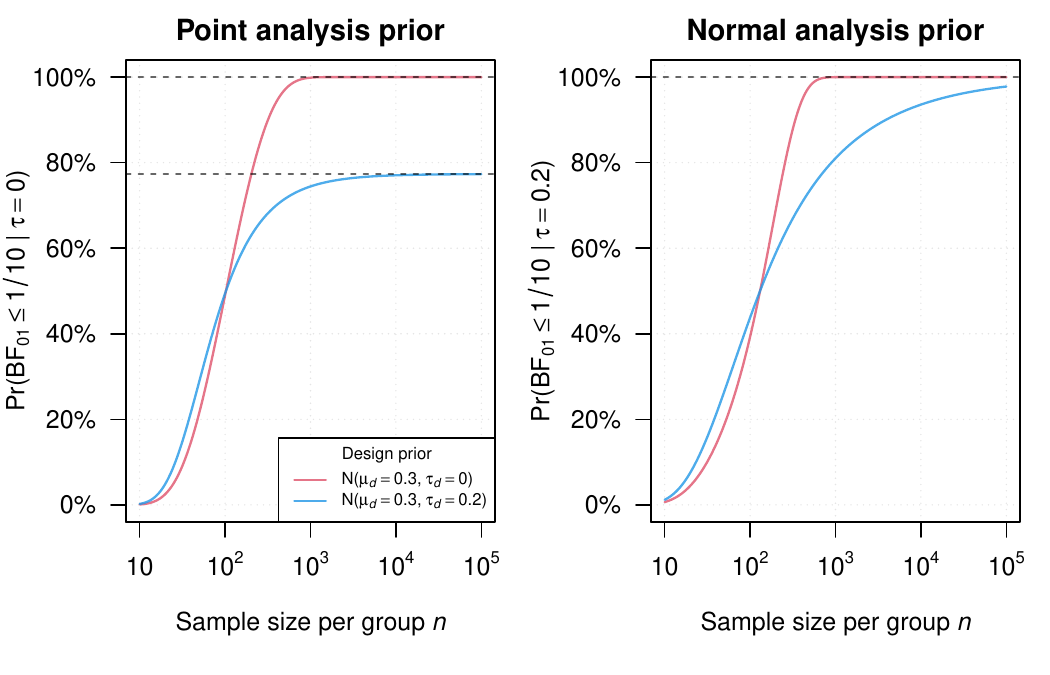} 
\end{knitrout}
\caption{Examples of Bayes factor power curves computed with
  equation~\eqref{eq:prLR} (left) and~\eqref{eq:prBF} (right). Both Bayes
  factors contrast $H_{0} \colon \theta = 0$ against
  $H_{1} \colon \theta \neq 0$ for a standardized mean difference
  parameter $\theta$ with unit variance
  $\sigma^{2}_{\scriptscriptstyle \hat{\theta}} = 2$ with either a point
  analysis prior ($\tau = 0$, left plot) or a normal analysis
  prior ($\tau = 0.2$, right plot) both with $\mu = 0.3$. The
  power is computed under either a point (red curve) or normal design prior
  (blue curve), both at the same location as the analysis prior. The dashed
  lines depict the corresponding limiting power values.}
\label{fig:powerexamples}
\end{figure}

The left plot in Figure~\ref{fig:powerexamples} shows examples of two power
functions related to the Bayes factor for a standardized mean difference
parameter with point analysis prior. The power is computed under either a point
(red curve) or a normal design prior (blue curve). We see that increasing the
sample size $n$ increases the power in both cases. However, as known
from~\eqref{eq:powerlimLR}, the power under the normal design prior is bounded
by
$1 - \Phi(Z_{\mathrm{lim}}) = 77.3\%$
while the power under the point design prior can be increased up to $100\%$.
Finally, the curves cross at 50\%, the intuition being that with the normal
design prior there is more uncertainty, and hence the power is always closer to
$50\%$.

For the Bayes factor with normal analysis prior ($\tau > 0$), the
cumulative distribution or power function is given by
\begin{align}
\label{eq:prBF}
    \Pr(\text{BF}_{01} \leq k \mid n, \mu_{d} , \tau_{d}, \tau > 0)
    = \Phi(-\sqrt{X} - M) + \Phi(-\sqrt{X} + M)
\end{align}
with 
\begin{align*}
    M = \left\{\mu_{d} - \theta_0 - \frac{\sigma^2_{\scriptscriptstyle \hat{\theta}}}{n\tau^2}(\theta_0 - \mu) \right\}\frac{1}{\sqrt{\tau^{2}_{d} + \sigma^{2}_{\scriptscriptstyle \hat{\theta}}/n}}
\end{align*}
and
\begin{align*}
    X = \left\{\log\left(1 + \frac{n\tau^2}{\sigma^2_{\scriptscriptstyle \hat{\theta}}}\right) + \frac{(\theta_0 - \mu)^2}{\tau^2} - \log k^2\right\} \left(1 + \frac{\sigma^2_{\scriptscriptstyle \hat{\theta}}}{n\tau^2} \right) \frac{\sigma^2_{\scriptscriptstyle \hat{\theta}}}{n\tau^{2}_{d} + \sigma^{2}_{\scriptscriptstyle \hat{\theta}}},
\end{align*}
see Appendix~\ref{app:distributions} for details. Again, to compute the
probability of a Bayes factor in favor of the null hypothesis, we have to take
one minus the probability~\eqref{eq:prBF} along with a level $k > 1$.

Unlike the power function
based on the Bayes factor with point analysis prior~\eqref{eq:prLR}, the power
function based on the Bayes factor with normal analysis prior~\eqref{eq:prBF} can
be increased arbitrarily close to one by increasing the sample size $n$ (see
Appendix~\ref{app:asymptotics}). That is, we have
\begin{align}
  \lim_{n\to \infty} \Pr(\text{BF}_{01} \leq k \mid n, \mu_{d} , \tau_{d}, \tau > 0) = 1
  \label{eq:plimBF}
\end{align}
regardless of whether the design prior is a point prior ($\tau_{d} = 0$) or
a normal prior ($\tau_{d} > 0$), provided that the design prior is not equal
to the point null hypothesis itself. This is expected because Bayes factors
contrasting point nulls against composite alternatives are `consistent' in the
sense that as the sample size increases, the probability of the Bayes factor
favoring the hypothesis under which the data were generated tends to one
\citep{Dawid2011,Bayarri2012,Ly2021}.

The right plot in Figure~\ref{fig:powerexamples} shows example power curves for
the Bayes factor related to a standardized mean difference parameter with normal
analysis prior, and computed under either point (red curve) or normal design
prior (blue curve). We see that in both cases an increase in sample size also
increases the power. As expected from~\eqref{eq:plimBF}, the power can increase
to $100\%$ in both cases, although it approaches the limit much slower under the
normal than under the point design prior because there is more uncertainty.
Finally, as with the point analysis prior Bayes factor, the curves cross at
50\%.

\section{Sample size determination}
\label{sec:ssd}
Both power functions~\eqref{eq:prLR} and~\eqref{eq:prBF} are straightforward to
implement and can be used to obtain power curves as a function of the sample
size, or of other parameters. We provide R implementations of both in our
package \texttt{bfpwr} (see Appendix~\ref{app:pkg} for an illustration).
Iterative root-finding can then be applied to determine the sample size such
that compelling evidence is obtained with a desired target power under a
specified design prior \citep[as pioneered by][]{Weiss1997}. It is important to
emphasize again that one can also compute a power curve in favor of the null
hypothesis. That is, one can look at one minus the power
functions~\eqref{eq:prLR} and~\eqref{eq:prBF} along with a Bayes factor
threshold $k > 1$ and fixing the design prior to the null hypothesis ($\mu_{d} =
\theta_{0}$, $\tau_{d} = 0$). In this way, sample sizes can be determined that
ensure a desired probability of compelling evidence for both the null and the
alternative.

We will now investigate situations where the sample size can be obtained in
closed-form. As for the distribution of the Bayes factor in the previous
section, there is again a distinction between sample size determination for
Bayes factors with point analysis priors ($\tau = 0$) and normal analysis priors
($\tau > 0$). We start again with the former.

\subsection{Bayes factor with point analysis prior}
Assuming that the alternative $\mu$ is larger than the null $\theta_{0}$ and
setting the power function~\eqref{eq:prLR} equal to a target power $1 - \beta$,
we obtain a quadratic equation in the sample size $n$.
Its solution can be expressed as
\begin{align}
\label{eq:nLR}
  n = \left[
  \left\{z_{1 - \beta} + \sqrt{z_{1 - \beta}^{2} - \frac{\Delta_{{\mu_{d}}}\log k^{2}}{\Delta_{\mu}} + \left(\frac{\tau_{d} \log k^{2}}{\Delta_{\mu}}\right)^{2}}\right\}^{2} - \left(\frac{\tau_{d} \log k^{2}}{\Delta_{\mu}}\right)^{2}\right]
  \times \frac{\sigma^2_{\scriptscriptstyle \hat{\theta}}}{
  \Delta_{\mu_{d}}^{2} - 4 z_{1 - \beta}^{2} \tau^{2}_{d}}
\end{align}
where $\Delta_{\mu_{d}} = 2\mu_{d} - \mu - \theta_{0}$ is the `generalized
effect size`, which reduces to the ordinary effect size
$\Delta_{\mu} = \mu - \theta_{0}$ when the design prior mean is set to the
parameter value under the alternative ($\mu_{d} = \mu$). From looking
at~\eqref{eq:nLR} one can recognize that a valid sample size can only exist if
the denominator in the right factor is positive, as sample sizes cannot be
negative. This condition is equivalent to the target power $1 - \beta$ being
lower than the limiting power~\eqref{eq:powerlimLR}. Of note, if for a given
design prior the limiting power~\eqref{eq:powerlimLR} is higher than 50\%,
replacing the first plus in~\eqref{eq:nLR} by a minus gives the sample size that
leads to a target power of $\beta$ instead of $1 - \beta$. It is also worth
noting that the sample size formula~\eqref{eq:nLR} will usually produce
non-integer values and hence needs to be rounded to the next larger integer in
order to be an evaluable sample size in practice (an actual number of
participants, animals, etc.).

To better understand the sample size formula~\eqref{eq:nLR}, we will now
investigate it closer for two special cases. First, suppose that the design
prior is a point prior ($\tau_{d} = 0$) at $\mu_{d}$, not necessarily the
same as the alternative $\mu$. This leads to~\eqref{eq:nLR} reducing to
\begin{align}
  \label{eq:nLR2}
  n = \frac{\sigma^2_{\scriptscriptstyle \hat{\theta}}
  \left\{z_{1 - \beta} + \sqrt{z_{1 - \beta}^2 -  (\Delta_{\mu_{d}}/\Delta_{\mu}) \log k^2}\right\}^2}{\Delta_{\mu_{d}}^2}.
\end{align}
Assuming that the tested parameter is a mean difference with unit variance
$\sigma^2_{\scriptscriptstyle \hat{\theta}} = 2\sigma^{2}$ (see
Table~\ref{tab:outcomes}, second row), we can see that~\eqref{eq:nLR2}
represents a modification of the frequentist sample size
formula~\eqref{eq:freqsamplesize}: The effect size
$\Delta_{\mu} = \mu - \theta_0$ in the denominator of the frequentist sample
size is replaced by the generalized effect size $\Delta_{\mu_{d}}$ that takes
into account the mean of the design prior $\mu_{d}$, but reduces to the effect
size when the design prior mean equals parameter under the alternative
($\mu_{d} = \mu$). Moreover, the quantile $z_{1-\alpha/2}$ in the frequentist
sample size is replaced by
$\surd (z^2_{1-\beta} - (\Delta_{\mu_{d}}/\Delta_{\mu}) \log k^2)$, reflecting
the fact that we are interested in a Bayes factor hypothesis test with evidence
threshold $k$ instead of a frequentist test with level $\alpha$.

Second, assume that the design prior is also equal to the alternative
($\mu_{d} = \mu$), so that the formula~\eqref{eq:nLR2} further reduces to
\begin{align}
\label{eq:nLR3}
    n = \frac{\sigma^2_{\scriptscriptstyle \hat{\theta}} \left\{z_{1 - \beta} + \sqrt{z_{1 - \beta}^2 - \log k^2}\right\}^2}{\Delta_{\mu}^2}.
\end{align}
The same formula~\eqref{eq:nLR3} was also found by \citet{Strug2007} for
`evidential' sample size calculations, but is unfortunately not well known. Not
surprisingly, the two formulas coincide, since Bayes factors and likelihood
ratios -- the measure of evidence used in evidential/likelihoodist statistics
\citep[see e.g.,][]{Edwards1971, Royall1997, Blume2002, Strug2018, Cahusac2020}
-- are equivalent when the Bayes factor involves only point hypotheses. Our more
general formulas~\eqref{eq:nLR} and~\eqref{eq:nLR2} thus enable `hybrid
Bayesian/likelihoodist' design that assumes a likelihoodist analysis but can
incorporate prior knowledge and uncertainty via Bayesian design prior, similar
to how Bayesian design priors can be used to incorporate parameter uncertainty
in the design of frequentist hypothesis tests \citep[see e.g.,][for an overview
of hybrid Bayesian/frequentist design approaches]{Grieve2022}.

\begin{table}[!htb]
\centering
\caption{Sample size per group $n$ to obtain a Bayes factor
  $\mathrm{BF}_{01} \leq k$ with at least a power of $1 - \beta$. The parameter
  of interest is a standardized mean difference and the analysis and design
  prior assume both an effect size of one ($\Delta_{\mu} = 1$) so that
  equation~\eqref{eq:nLR3} can be used to compute the sample size in
  closed-form.}
\label{tab:smdn}

  \rowcolors{1}{}{lightgray}
\begin{tabular}{rrrrrrrrrrrrr}
  \toprule
  & \multicolumn{12}{c}{$k$} \\
 \cmidrule{2-13} 
 $1 - \beta$ & 1/3 & 1/4 & 1/5 & 1/6 & 1/7 & 1/8 & 1/9 & 1/10 & 1/30 & 1/100 & 1/300 & 1/1000 \\ \midrule
50\% & 5 & 6 & 7 & 8 & 8 & 9 & 9 & 10 & 14 & 19 & 23 & 28 \\ 
  55\% & 6 & 7 & 8 & 9 & 9 & 10 & 10 & 11 & 15 & 21 & 25 & 30 \\ 
  60\% & 7 & 8 & 9 & 10 & 11 & 11 & 12 & 12 & 17 & 22 & 27 & 32 \\ 
  65\% & 8 & 9 & 10 & 11 & 12 & 13 & 13 & 14 & 19 & 24 & 29 & 34 \\ 
  70\% & 9 & 11 & 12 & 13 & 14 & 14 & 15 & 15 & 21 & 26 & 32 & 37 \\ 
  75\% & 11 & 13 & 14 & 15 & 16 & 16 & 17 & 18 & 23 & 29 & 34 & 40 \\ 
  80\% & 13 & 15 & 16 & 17 & 18 & 19 & 20 & 20 & 26 & 32 & 38 & 44 \\ 
  85\% & 17 & 18 & 20 & 21 & 22 & 23 & 23 & 24 & 30 & 37 & 42 & 48 \\ 
  90\% & 22 & 23 & 25 & 26 & 27 & 28 & 28 & 29 & 36 & 42 & 48 & 55 \\ 
  95\% & 30 & 32 & 34 & 35 & 36 & 37 & 38 & 38 & 45 & 52 & 59 & 66 \\ 
   \bottomrule
\end{tabular}

\end{table}

To illustrate formula~\eqref{eq:nLR3}, we now assume that $\hat{\theta}$ is a
standardized mean difference so that the unit variance is
$\sigma^2_{\scriptscriptstyle \hat{\theta}} = 2$ and $n$ can be interpreted as
the sample size per group (see the third row in Table~\ref{tab:outcomes}).
Table~\ref{tab:smdn} shows the sample size~\eqref{eq:nLR3} based on an assumed
effect size of one ($\Delta_{\mu} = 1$). We see, for instance, that a sample
size of
$n = 20$
per group is required to achieve $1 - \beta = 80\%$ power for a Bayes factor
threshold $k = 1/10$. If the assumed effect size was smaller, the required
sample size would become larger. For example, for a half as large effect size,
the sample sizes from Table~\ref{tab:smdn} quadruple, e.g., requiring a sample
size of
$n = 80$
per group to have $1 - \beta = 80\%$ power for a Bayes factor threshold of
$k = 1/10$.

\begin{figure}[!htb]
\begin{knitrout}
\definecolor{shadecolor}{rgb}{0.969, 0.969, 0.969}\color{fgcolor}
\includegraphics[width=\maxwidth]{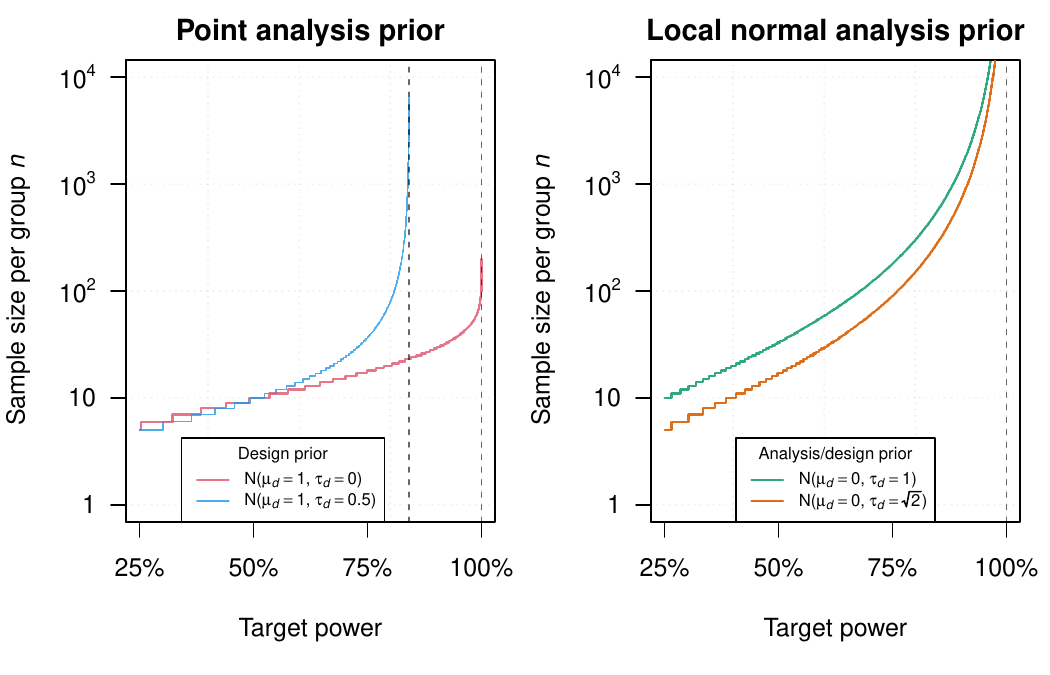} 
\end{knitrout}
\caption{Examples of Bayes factor sample size calculations with
  equation~\eqref{eq:nLR} (left) and equation~\eqref{eq:nBF} (right). Both Bayes
  factors contrast $H_{0} \colon \theta = 0$ against
  $H_{1} \colon \theta \neq 0$ for a standardized mean difference
  parameter $\theta$ with unit variance
  $\sigma^{2}_{\scriptscriptstyle \hat{\theta}} = 2$ with either a point
  analysis prior ($\mu = 1$, $\tau = 0$, left plot) or a
  local normal analysis prior ($\mu = 0$, right plot), and at a
  threshold of $k = 1/10$. The dashed lines represent the upper bounds
  for the power that are achievable with a finite sample size.}
\label{fig:nexamples}
\end{figure}

The left plot in Figure~\ref{fig:nexamples} shows sample size calculations for a
Bayes factor at threshold $k = 1/10$ with the same point analysis and
design prior as in Table~\ref{tab:smdn} (red curve), but additionally
illustrates the sample size formula~\eqref{eq:nLR} that also incorporates
parameter uncertainty via a normal design prior (blue curve). We can see that
for a target power above 50\%, larger sample sizes are required under the normal
design prior than under the point design prior, while it is reversed for a
target power below 50\%. Furthermore, we see that as the target power approaches
its theoretical upper bound~\eqref{eq:powerlimLR}, the sample size goes to
infinity.

\subsection{Bayes factor with normal analysis prior}
Finding a sample size formula becomes more difficult when we move from point to
normal analysis priors. The technical reason is that when we set the power
function~\eqref{eq:prBF} equal to a target power and try to solve for the sample
size $n$, we have $n$ appearing both in and outside logarithms. This forms a
transcendental equation that cannot be solved in terms of elementary functions.
We were unable to find a closed-form solution in the general case. However, as
we will show now, solutions can, under certain conditions, be expressed in terms
of the Lambert W function \citep{Corless1996}. The Lambert W function is the
function $\mathrm{W}(\cdot)$ that satisfies
$\mathrm{W}(x)\exp\{\mathrm{W}(x)\}=x$, and it is therefore sometimes also
called `product logarithm'. It has many fundamental applications and has also
previously appeared in the context of Bayes factor hypothesis testing
\citep{Pawel2022b, Wagenmakers2022, Held2021b, Pawel2023}.

Suppose now that the design and analysis prior are both centered around the null
value \mbox{($\mu_{d} = \mu = \theta_0$)}. Centering the prior around the null
value is commonly done in `default' Bayes factor tests \citep{Berger1987b}. It
encodes the assumption that some parameters are larger while others are smaller
than the null, the standard deviation of the distribution determining the
variability, yet the average parameter equals the null value. We then have that
$M=0$ and hence the power~\eqref{eq:prBF} reduces to
\begin{align}
  \label{eq:prBFcenter}
    \Pr\{\text{BF}_{01} \leq k \mid n, \tau_{d}, \mu_{d} = \mu = \theta_{0}\}
    = 2\Phi(-\sqrt{X}).
\end{align}
Further, assume that the variance of the design prior corresponds to the
variance of the analysis prior ($\tau_{d} = \tau$). We then have that
\begin{align*}
    X = \left\{\log\left(1 + \frac{n\tau^2}{\sigma^2_{\scriptscriptstyle \hat{\theta}}}\right) - \log k^2\right\} \frac{\sigma^2_{\scriptscriptstyle \hat{\theta}}}{n\tau^2}.
\end{align*}
Setting the power function~\eqref{eq:prBFcenter} equal to a target power of
$1 - \beta$ and assuming that
$\log\{1 + (n\tau^2)/\sigma^2_{\scriptscriptstyle \hat{\theta}}\} \approx \log\{(n\tau^2)/\sigma^2_{\scriptscriptstyle \hat{\theta}}\}$,
we obtain the following approximate sample size formula
\begin{align}
\label{eq:nBF}
  n = \frac{\sigma^2_{\scriptscriptstyle \hat{\theta}}}{\tau^2} \, \underbrace{k^2  \exp\left\{-  \mathrm{W}_{-1}(-k^2 \, z^2_{(1 - \beta)/2})\right\}}_{=n_{k,\beta}}
\end{align}
with $\mathrm{W}_{-1}(\cdot)$ the branch of the Lambert W function that
satisfies $\mathrm{W}(x) < -1$ for $y \in (-1/e, 0)$, see
Appendix~\ref{app:lambertWderiv} for details.

We can see that the sample size~\eqref{eq:nBF} depends on the ratio of the unit
variance $\sigma^{2}_{\scriptscriptstyle \hat{\theta}}$ to the prior variance
$\tau^{2}$ multiplied by a `unit information sample size' $n_{k,\beta}$ which
depends only on the Bayes factor threshold $k$ and the target power $1 - \beta$.
The unit information sample size is the sample size that is obtained when a unit
information prior \citep{Kass1995b} is specified, which is a prior with variance
equal to the unit variance
($\tau^{2} = \sigma^{2}_{\scriptscriptstyle \hat{\theta}}$). As with the
frequentist sample size~\eqref{eq:freqsamplesize}, smaller unit variances
$\sigma^{2}_{\scriptscriptstyle \hat{\theta}}$ reduce the sample
size~\eqref{eq:nBF}. The prior variance $\tau^{2}$ determines how large
parameters are expected under the alternative hypothesis, and as such, larger
prior variances lead to a reduction of sample size similar to how larger effect
sizes lead to a reduction of sample size in the frequentist
formula~\eqref{eq:freqsamplesize}. Finally, the formula~\eqref{eq:nBF} allows us
to study the potential existence of a sample size that can achieve the target
power: Since the argument of the Lambert W function has to be at least $-1/e$
for it to be defined, we can infer that only combinations of Bayes factor
thresholds $k$ and power values $1 - \beta$ that satisfy
$-k^{2} z^{2}_{(1 - \beta)/2} \geq -1/e$ can actually be achievable with a finite
sample size. For example, it is impossible to find a sample size that guarantees
a power of $1 - \beta = 50\%$ for a threshold of $k = 1$ since then
$- 1 \times z_{0.25}^{2} = -0.45 < -1/e = -0.37$.

\begin{table}[!htb]
\centering
\caption{Required unit information sample size $n_{k,\beta}$ computed with
  equation~\eqref{eq:nBF} to obtain a Bayes factor $\mathrm{BF}_{01} \leq k$
  with at least a power of $1-\beta$ with a unit information analysis and design
  prior.}
\label{tab:nnormal}

  \rowcolors{1}{}{lightgray}
\begin{tabular}{rrrrrrrrrrrrr}
  \toprule
  & \multicolumn{12}{c}{$k$} \\
 \cmidrule{2-13} 
 $1 - \beta$ & 1/3 & 1/4 & 1/5 & 1/6 & 1/7 & 1/8 & 1/9 & 1/10 & 1/30 & 1/100 & 1/300 & 1/1000 \\ \midrule
50\% & 10 & 12 & 13 & 14 & 15 & 16 & 16 & 17 & 22 & 28 & 33 & 39 \\ 
  55\% & 14 & 16 & 17 & 19 & 20 & 21 & 21 & 22 & 29 & 36 & 43 & 50 \\ 
  60\% & 19 & 22 & 24 & 25 & 27 & 28 & 29 & 29 & 38 & 48 & 57 & 66 \\ 
  65\% & 27 & 30 & 33 & 35 & 37 & 38 & 40 & 41 & 53 & 66 & 77 & 89 \\ 
  70\% & 40 & 45 & 48 & 51 & 53 & 56 & 57 & 59 & 75 & 93 & 109 & 126 \\ 
  75\% & 63 & 70 & 75 & 79 & 82 & 85 & 88 & 90 & 114 & 140 & 163 & 188 \\ 
  80\% & 108 & 118 & 126 & 132 & 138 & 143 & 147 & 150 & 188 & 229 & 265 & 305 \\ 
  85\% & 212 & 230 & 244 & 256 & 265 & 274 & 281 & 287 & 355 & 427 & 493 & 564 \\ 
  90\% & 538 & 579 & 610 & 636 & 658 & 677 & 693 & 708 & 859 & 1023 & 1170 & 1331 \\ 
  95\% & 2554 & 2716 & 2841 & 2943 & 3029 & 3103 & 3168 & 3226 & 3829 & 4481 & 5071 & 5714 \\ 
   \bottomrule
\end{tabular}

\end{table}

Table~\ref{tab:nnormal} shows unit information sample sizes $n_{k,\beta}$ for a
range of powers $1 - \beta$ and Bayes factor thresholds $k$. Compared to the
sample sizes from Table~\ref{tab:smdn} the sample sizes are quite a bit larger.
This is because the design and analysis priors underlying each of these
calculations encode vastly different assumptions: The local normal analysis
prior from Table~\ref{tab:nnormal} represents a parameter distribution that is
centered around the null value while the point analysis prior from
Table~\ref{tab:smdn} represents a mean difference of one standard deviation away
from the null. The former with unit information variance represents a more
pessimistic assumption about the parameter than the latter. To incorporate more
optimistic beliefs into the calculations we may increase the standard deviation
$\tau$ of the distribution as this encodes the assumption of potentially larger
parameters. For example, doubling $\tau$ leads to a four-fold decrease of the
sample size~\eqref{eq:nBF}. This is also illustrated in the right plot of
Figure~\ref{fig:nexamples} where the prior with doubled variance (orange curve)
leads to a two-fold decrease in sample size over the prior with variance of one
(green curve).

The formula~\eqref{eq:nBF} is, to our knowledge, the first closed-form sample
size formula for Bayes factor analysis with normal analysis priors. While
interesting from a theoretical point of view, its practical use is perhaps more
limited than the sample size formula for Bayes factors with point analysis
priors~\eqref{eq:nLR}. This is because it makes the restrictive assumption that
the design prior and the analysis prior are both centered around the null
($\mu_{d} = \mu = \theta_{0}$). This seems unrealistic in practice, since
researchers designing a study usually have good reasons to expect parameters to
be different from the null and would like to account for this in the sample size
calculation. However, given our limited mathematical abilities, we have not been
able to derive a sample size formula for this more general setting due to the
transcendental nature of the power equation. Fortunately, the sample size can
still be easily calculated numerically with our R package which is quicker and
more reliable than computing it with a simulation approach.

\section{Application}
\label{sec:application}
We will now illustrate Bayes factor sample size and power calculations using
examples from medicine and psychology.

\subsection{Randomized controlled clinical trial on mirtazapine in dementia}
\citet{Banerjee2021} conducted a randomized controlled clinical trial to assess
the effect of the antidepressant mirtazapine on agitated behavior in patients
with dementia. The primary outcome of the trial was the Cohen-Mansfield
Agitation Inventory (CMAI) score at 12 weeks after treatment start. The CMAI
score is commonly used to quantify agitation in dementia and it can take values
from 29 to 203. The treatment effect was quantified as the difference in mean
CMAI between the mirtazapine and placebo groups $\theta$. The null hypothesis
was defined as no mean difference between mirtazapine and placebo ($H_{0} \colon
\theta = 0$), while the alternative hypothesis was formulated as a mean decrease
of six CMAI points ($H_{1} \colon \theta = -6$). With a target power of
$1 - \beta = 80\%$, a two-sided level of $\alpha = 0.05$,
and an assumed standard deviation of $\sigma = 15$ CMAI points,
formula~\eqref{eq:freqsamplesize} gives a required per-group sample size of $n =
99$, which the investigators adjusted further upwards to 111 to
account for attrition. Contrary to their expectations, however, the trial did
not find a statistically significant effect of mirtazapine over placebo
(estimated mean difference of $\hat{\theta} = -1.74$ with 95\% confidence
interval from $-7.17$ to $3.69$ and two-sided $p$-value of $p =
0.53$ for the null hypothesis of no difference).

While this study was designed and analyzed using frequentist methodology, we
will now examine what the analysis, power, and sample size calculations might
have looked like if the planned analysis were performed using Bayes factors. To
do this, we assume a point analysis prior equal to the alternative hypothesis
specified by the trial investigators ($\mu = -6$ and $\tau =
0$), so that the Bayes factor corresponds to a likelihood ratio. Based on the
estimated mean difference $\hat{\theta} = -1.74$ and its
standard error $\sigma_{\hat{\theta}}/\sqrt{n} = 2.77$
(recalculated from the confidence interval), we can apply
equation~\eqref{eq:BF01} and obtain a Bayes factor $\text{BF}_{01} =
2.7$. This Bayes factor indicates only anecdotal evidence in
favor of the null hypothesis of no effect of mirtazapine over the specified
alternative.

\begin{figure}[!tb]
\begin{knitrout}
\definecolor{shadecolor}{rgb}{0.969, 0.969, 0.969}\color{fgcolor}
\includegraphics[width=\maxwidth]{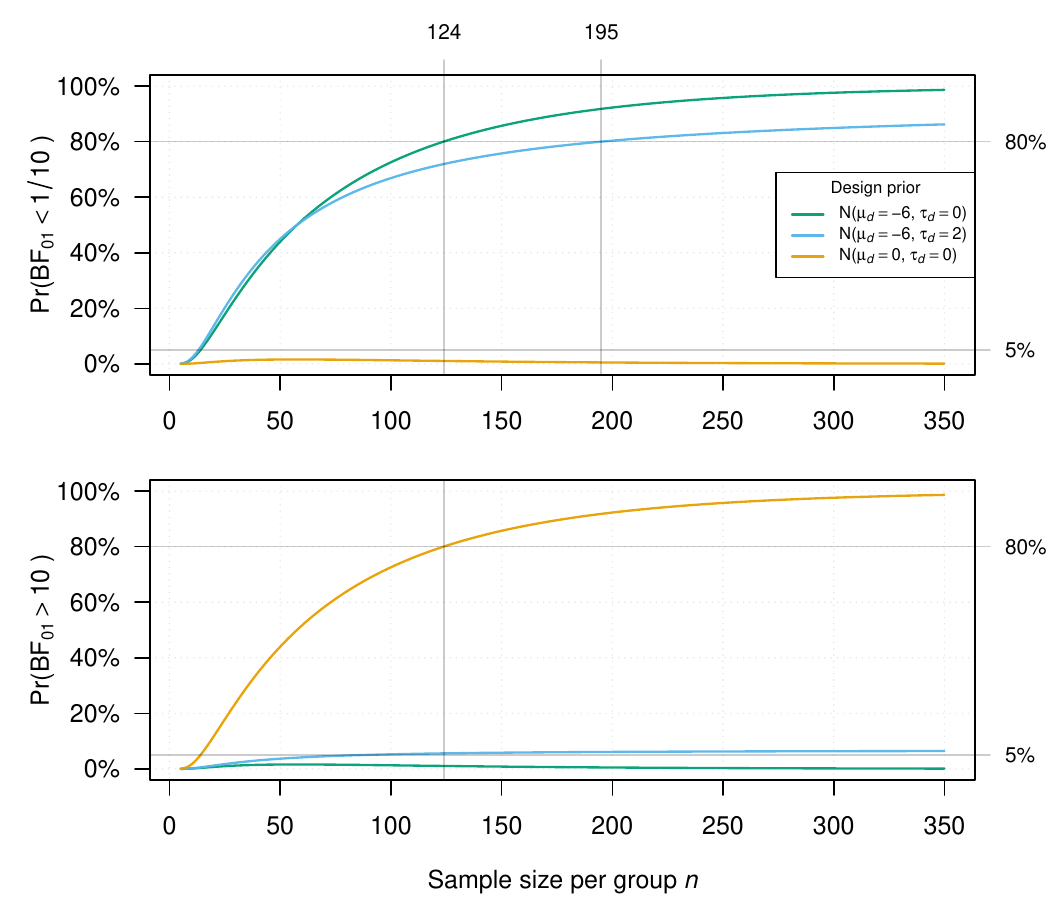} 
\end{knitrout}
\caption{Power and sample size calculations for randomized clinical trial
  assessing the effect of mirtazapine on agitated behavior in patients with
  dementia \citep{Banerjee2021}. The Bayes factor assumed for the analysis
  contrasts $H_{0} \colon \theta = 0$ to $H_{1} \colon \theta =
  -6$ related to the mean difference $\theta$ in CMAI scores between
  mirtazapine and placebo. The power is computed under either a point prior at
  the assumed effect under the alternative (green), a normal prior that
  incorporates additional parameter uncertainty (blue), or under the null
  hypothesis (orange).}
\label{fig:radiotherapy}
\end{figure}

Taking a step back and assuming that no data have yet been collected, we can use
equations~\eqref{eq:prLR} and \eqref{eq:nLR} to calculate power and sample size.
Figure~\ref{fig:radiotherapy} shows power curves based on the Bayes factor
providing strong evidence in favor of the alternative ($\mathrm{BF}_{01} <
1/10$) in the top plot, and based on the Bayes factor providing strong
evidence in favor of the null ($\mathrm{BF}_{01} > 10$) in the bottom
plot. The colors indicate under which design prior the power was computed.

Focusing on the top plot, we can see from the green curve that at least $n =
124$ observations per group are required to ensure a target power of
$1 - \beta = 80\%$ assuming that the same point prior is
used in the design as for the analysis (i.e., a point prior at $\mu =
-6$). This number increases to $n = 195$ when we move to a
design prior that incorporates parameter uncertainty (blue curve), i.e., a prior
that is still centered around $\mu = -6$ but with a standard deviation
of $\tau_{d} = 2$. Finally, if we look at the orange power curve
computed assuming that the null hypothesis is true ($\theta = 0$), we
can see that the probability of misleading evidence for the alternative when the
null hypothesis is actually true (the type I error rate) is very low and appears
to be reasonably controlled by conventional standards (i.e., below 5\%) for each
of the two samples sizes. This curve could be presented to a trial regulator to
demonstrate that the design is adequately calibrated.

Focusing now on the bottom plot, we can see that the sample size $n =
124$ based on the point prior also ensures a power of $1 - \beta =
80\%$ for finding evidence for the null hypothesis. This
is due to the symmetric nature of the point versus point hypothesis Bayes factor
considered in this example, as swapping the null $\theta_{0}$ and alternative
$\mu$ in formula~\eqref{eq:nLR3} does not change the resulting sample size.
Finally, looking at the green and blue curves we can see that the probability of
misleading evidence in favor of the null when the data are generated from the
non-null design priors seems to be reasonably well controlled (below the
conventional 5\%) for the point design prior (green), while it is slightly
inflated for the normal design prior (blue) for sample sizes larger than
88.

\subsection{Comparison to Monte Carlo simulation methods}
\label{sec:schoenbrodtexample}
\citet{Schoenbrodt2017} proposed a Monte Carlo simulation approach for power and
sample size calculations for Bayes factor analyses, and provide the R package
\texttt{BFDA} \citep{Schoenbrodt2019} for this purpose. The idea is to simulate
data sets under an assumed design prior and sample size, and then analyze each
data set with a specified Bayes factor. This results in a distribution of Bayes
factors from which the power can be calculated. The simulation is then repeated
for other sample sizes until the desired power is achieved. This approach can be
used in quite general settings, but can be computationally intensive and comes
with Monte Carlo error.

\begin{figure}[!tb]
\begin{knitrout}
\definecolor{shadecolor}{rgb}{0.969, 0.969, 0.969}\color{fgcolor}
\includegraphics[width=\maxwidth]{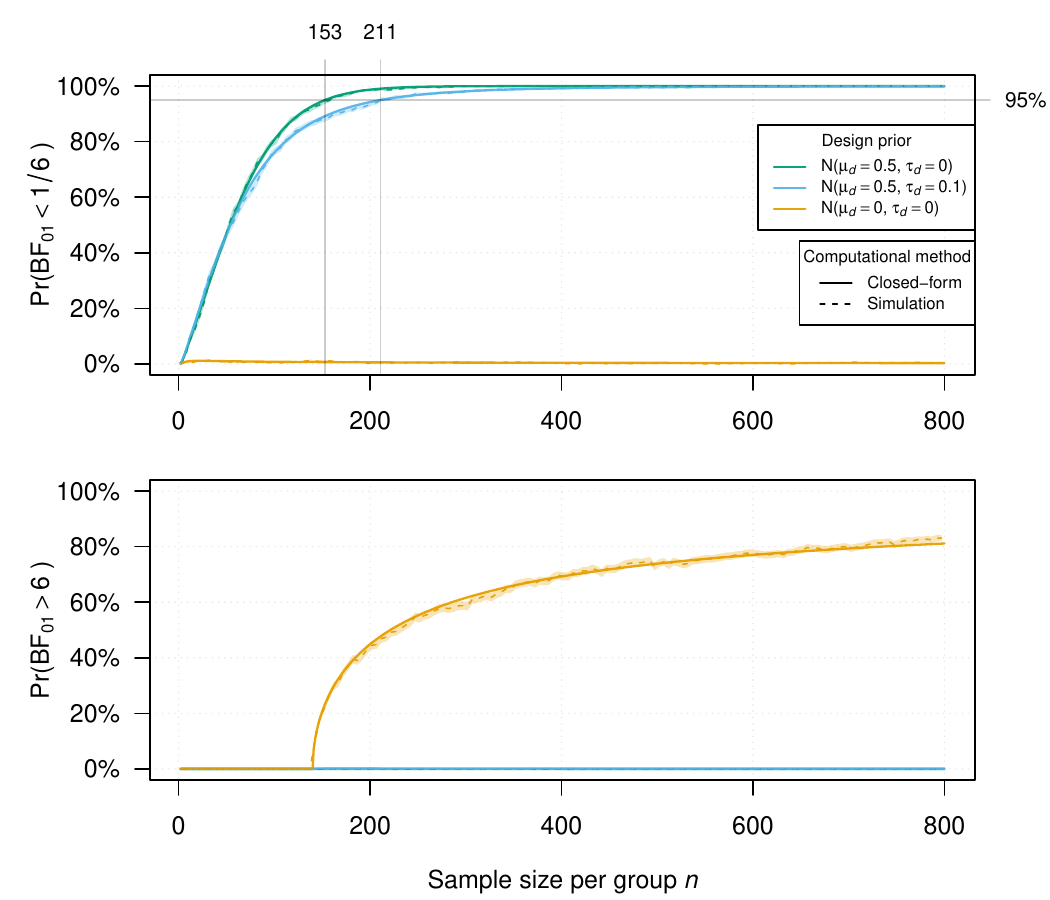} 
\end{knitrout}
\caption{Power and sample size calculation for standardized mean difference
  example. The Bayes factor assumed for the analysis contrasts $H_{0} \colon
  \theta = 0$ to $H_{1} \colon \theta \neq 0$ with
  \mbox{$\theta \mid H_{1} \sim \mathrm{N}(\mu = 0, \tau^{2} =
    1/2)$} prior assigned to the standardized mean difference
  $\theta$ under the alternative. The power is computed under either a point
  prior at a medium effect size (green), a normal prior that incorporates
  additional parameter uncertainty (blue), or under the null hypothesis
  (orange). For comparison, 1000 Monte Carlo simulations are performed
  with the \texttt{BFDA} package to obtain a simulation-based power curve
  approximation (dashed line with one Monte Carlo standard error band).}
\label{fig:comparisonBFDA}
\end{figure}

We will now look at an adaptation of an example examined in
\citet[p.~133]{Schoenbrodt2017}. Suppose we want to test the null hypothesis
that a standardized mean difference $\theta$ is zero ($H_{0} \colon \theta =
0$) versus the alternative hypothesis that it is different from zero
($H_{1} \colon \theta \neq 0$). We assume a $\theta \mid H_{1} \sim
\mathrm{N}(0, 1/2)$ analysis prior for the standardized
mean difference under the alternative, similar to the Cauchy prior with scale
$1/\sqrt{2}$ that was assumed by Schönbrodt and Wagenmakers. As
they did, we also investigate two design priors: either a point prior at
$\mu_{d} = 0.5$ , which is a convention for a `medium` standardized mean
difference in psychology \citep{Cohen1992}, or a normal prior at $\mu_{d} =
0.5$ with standard deviation $\tau_{d} = 0.1$ to incorporate
parameter uncertainty. Figure~\ref{fig:comparisonBFDA} shows the resulting power
curves computed from equation~\eqref{eq:prBF}, along with the \texttt{BFDA}
Monte Carlo simulation approximations (with Monte Carlo standard error bands)
for comparison.

We see that closed-form and simulation curves closely align in all cases. The
latter shows a maximum Monte Carlo error of 0.016. This Monte Carlo error may or may not be negligible for
practical purposes, as it may lead to slight over- or underestimation of the
sample size. For example, in animal studies, an overestimation of the sample
size by even one animal may be highly problematic and thus a higher number of
simulations may be preferred, whereas in other studies this would not be a
problem at all.

Looking at the green curves in the top plot, we see that a sample size of $n =
153$ per group is sufficient to achieve a target power of
$95\%$ with a Bayes factor threshold of $k = 1/6$ under
the point design prior. This sample size is slightly larger than the $n = 146$
that Schönbrodt and Wagenmakers obtained in their calculations, which instead
assumed a Cauchy analysis prior. As in the previous example, incorporating
parameter uncertainty via a normal design prior increases the required sample
size, in this case to $n = 211$. Looking at the orange curve in the
bottom plot, we see that these sample sizes only have modest power of around
$20\%$ and $50\%$, respectively, for obtaining evidence in favor of the true
null hypothesis (at a threshold of $k = 6$). To obtain a power of
$95\%$, a sample size of $n = 6691$ per group would be
required (not shown in Figure~\ref{fig:comparisonBFDA}). This illustrates the
well-known fact that evidence accumulates slower for the null than the
alternative when the Bayes factor involves testing a point null against a
composite alternative with normal analysis prior \citep{Johnson2010}. Finally,
looking at the orange curve in the top plot and the green/blue curves in the
bottom plot, we see that the probability of misleading evidence in favor of the
incorrect hypothesis appears to be adequately controlled, as these curves are
virtually constant at zero over the entire range of sample sizes.

Appendix~\ref{app:simeval} shows a more systematic comparison with Monte Carlo
simulation, including many other design and analysis prior conditions. To verify
that the methods work as intended under a given condition, a sample size was
first calculated using closed-form (if available) or root-finding methods to
ensure a desired target power for a specified Bayes factor threshold.
Subsequently, data were simulated based on this sample size and the design
prior. Bayes factors were then calculated from the simulated data, and power was
estimated empirically from the proportion of Bayes factors below the specified
threshold. These simulation-based power estimates were spread closely around the
specified target power in all conditions.

\section{Extensions}
\label{sec:extensions}
The types of Bayes factors that we considered so far are limited to data in the
form of asymptotically normally distributed parameter estimates, and normal or
point priors in the analysis. Researchers may also want to use different data
models or prior distributions in the analysis. In the following, we outline two
extensions that modify the Bayes factor used for the analysis, while still
retaining the normal likelihood and point/normal design prior used for the
design.

\subsection{Informed Bayesian \textit{t}-test}

\citet{Gronau2020} proposed a Bayes factor for testing a standardized mean
difference parameter based on normally distributed data with unknown variance --
the same situation where a classical $t$-test would be used. The Bayes factor is
given by
\begin{align}
  \mathrm{BF}_{01} = \frac{\mathrm{T}_{\nu}(t \mid 0, 1)}{\int_{-\infty}^{+\infty}
  \mathrm{NCT}_{\nu}(t \mid \theta \sqrt{n}) \, \mathrm{T}_{\kappa}(\theta \mid \mu, \tau)_{[a,b]}
  \, \mathrm{d}\theta}
  \label{eq:tBF}
\end{align}
with $t$ the $t$-test statistic, $n$ the effective sample size (the actual
number of observations/pairs for one-sample/paired $t$-tests, or
$n = 1/(1/n_{1} + 1/n_{2})$ for two-sample $t$-tests),
$\mathrm{T}_{\nu}(\cdot \mid \mu, \tau)$ the location-scale $t$ density with
degrees of freedom $\nu$, location $\mu$, and scale $\tau$, and
$\mathrm{NCT}_{\nu}(\cdot \mid \lambda)$ the non-central $t$ density with
degrees of freedom $\nu$ and non-centrality parameter $\lambda$ \citep[chapters
28 and 31]{Johnson1995}. The subscript $[a,b]$ denotes truncation of a
distribution to the interval from $a$ to $b$, i.e.,
$f(x)_{[a,b]} = \{f(x) 1_{[a,b]}(x)\}/\{F(b) - F(a)\}$. For example, for a
one-sided test in positive direction, we have $a = 0$ and $b=+\infty$, while for
a two-sided test we have $a = -\infty$ and $b=+\infty$. When the prior under the
alternative is set to a (scaled) Cauchy distribution ($\kappa = 1$, $\mu = 0$,
$a = -\infty$, $b=+\infty$), the Bayes factor reduces to the
`Jeffreys-Zellner-Siow' (JZS) Bayes factor \citep{Jeffreys1939, Zellner1980},
which is often used as a `default' Bayes factor in the social sciences
\citep{Rouder2009}. Setting other values for these parameters allows data
analysts to incorporate directionality, such as a one-sided JZS Bayes factor
\citep{Wetzels2009}, or prior knowledge about the standardized mean difference,
potentially improving the efficiency of the test \citep{Stefan2019}. The Bayes
factor~\eqref{eq:tBF} is also implemented in our package in the function
\texttt{tbf01}.

A complication in power and sample size calculations based on the Bayes
factor~\eqref{eq:tBF} is that it is not available in closed-form but requires
numerical evaluation of the integral in the denominator. Perhaps for this
reason, so far power and sample size have been computed using simulation. To
compute the power for a given sample size without simulation, we propose to use
the following two-step procedure instead:

\begin{enumerate}
  \item For a given sample size $n$ and analysis prior (specified with the
        parameters $\mu, \tau, \kappa, a, b$), use numerical root-finding to
        determine the `success region' $S$ in terms of $t$-statistics where the
        Bayes factor is equal or below the threshold $k$, i.e.,
        $S = \{t : \mathrm{BF}_{01} \leq k\}$. For a two-sided test, $S$ is
        typically a region specified by two critical values
        \mbox{$S = (-\infty, t_{\mathrm{crit}-}] \cup [t_{\mathrm{crit}+}, +\infty)$}.

  \item For a given design prior, compute the probability of obtaining a
        $t$-statistic included in $S$. For example, based on a normal design
        prior $\theta \sim \mathrm{N}(\mu_{d}, \tau^{2}_{d})$ and assuming a
        two-sample test with equally sized groups and that the data variance
        $\sigma^{2}$ is known, we have that approximately
        \mbox{$t \mid n, \mu_{d}, \tau_{d} \sim \mathrm{N}\{\mu_{d}\sqrt{n/2}, 1 + (n\tau^{2})/2\}$},
        hence the power can be computed by
  \begin{align*}
    \Pr\left(\mathrm{BF}_{01} \leq k \mid n, \mu_{d}, \tau_{d} \right)
    = \Phi\left(\frac{t_{\mathrm{crit}-} - \mu_{d}\sqrt{n/2}}{\sqrt{1 + (n\tau^{2}_{d})/2}}\right)
    + \Phi\left(\frac{\mu_{d}\sqrt{n/2} - t_{\mathrm{crit}+}}{\sqrt{1 + (n\tau^{2}_{d})/2}}\right).
  \end{align*}
\end{enumerate}

To calculate a sample size that ensures a certain target power, we can again use
numerical root-finding. To illustrate this, we now revisit the example from
\citet{Schoenbrodt2017} which was already re-analyzed in
Section~\ref{sec:schoenbrodtexample} with a normal analysis prior. We now
specify the analysis parameters $\mu = 0, \tau = 1/\sqrt{2}, \kappa = 1, a = 0,
b = +\infty$, i.e., a one-sided JZS Bayes factor with scale $1/\sqrt{2}$
\citep{Wetzels2009}, as in the original analysis. Using the previously described
iterative approach, we calculate that a sample size of \mbox{$n =
  143$} is required under a point design prior at $\mu_{d} =
0.5$ to achieve a target power of 95\%. This
differs only slightly from the simulation-based $n = 146$ reported by Schönbrodt
and Wagenmakers, possibly due to Monte Carlo error.

\subsection{Normal moment priors}
\begin{figure}[!htb]
\begin{knitrout}
\definecolor{shadecolor}{rgb}{0.969, 0.969, 0.969}\color{fgcolor}
\includegraphics[width=\maxwidth]{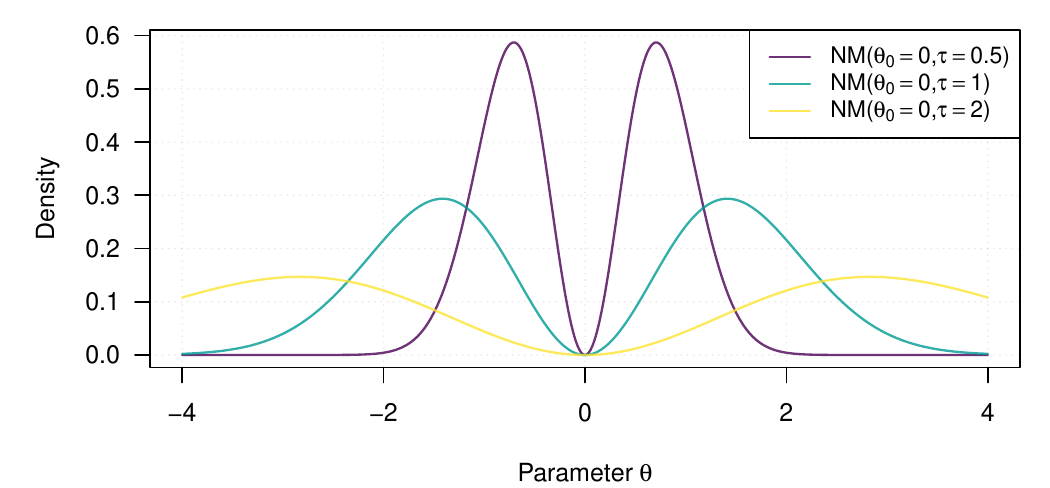} 
\end{knitrout}
\caption{Illustration of a normal moment prior distribution for different spread
  parameters $\tau$. }
\label{fig:nonlocal}
\end{figure}

Non-local priors are prior distributions that have zero probability density/mass
at the null value. They were introduced to allow evidence for the null to
accumulate more quickly if it is indeed true \citep{Johnson2010}. A convenient
class of non-local priors are normal moment priors, which have a density of the
form $\mathrm{NM}(\theta \mid \theta_{0}, \tau) = \mathrm{N}(\theta \mid
\theta_{0}, \tau^{2}) \times (\theta - \theta_{0})^{2}/\tau^{2}$ with location
$\theta_{0}$ and spread $\tau$. Figure~\ref{fig:nonlocal} shows three examples
of normal moment priors. We see that the distributions have a density of zero at
the null value $\theta_{0} = 0$, encoding the assumption that the null is the
least likely parameter value under the alternative. Normal moment priors have
two modes at $\pm \tau \sqrt{2}$, which gives a convenient way to elicit a
prior, as $\tau$ may, for example, be set so that the modes equals two parameter
values deemed most plausible under the alternative.

We will now extend power and sample size calculation to Bayes factors with
non-local moment analysis priors and normal design priors. The Bayes factor based
on a normally distributed parameter estimate
$\hat{\theta} \mid \theta \sim \mathrm{N}(\theta, \sigma^{2}_{\scriptscriptstyle \hat{\theta}}/n)$
that contrasts $H_{0} \colon \theta = \theta_{0}$ to
$H_{1} \colon \theta \neq \theta_{0}$ with prior
$\theta \mid H_{1} \sim \mathrm{NM}(\theta_{0}, \tau)$ assigned to $\theta$
under $H_{1}$ is given by
\begin{align}
  \label{eq:nlBF}
  \mathrm{BF}_{01} =
  \left(1 + \frac{n \tau^{2}}{\sigma^{2}_{\scriptscriptstyle \hat{\theta}}}\right)^{3/2}
  \, \exp\left[-\frac{1}{2} \frac{n(\hat{\theta} - \theta_{0})^{2}}{
  \sigma^{2}_{\scriptscriptstyle \hat{\theta}}\{1 + \sigma^{2}_{\scriptscriptstyle \hat{\theta}}/(n \tau^{2})\}}\right] \,
  \left[1 + \frac{n(\hat{\theta} - \theta_{0})^{2}}{
  \sigma^{2}_{\scriptscriptstyle \hat{\theta}}\{1 + \sigma^{2}_{\scriptscriptstyle \hat{\theta}}/(n \tau^{2})\}}\right]^{-1}
\end{align}
see e.g., \citet{Pramanik2024} or \citet{Pawel2023}. Under a normal design prior
\mbox{$\theta \sim \mathrm{N}(\mu_{d}, \tau^{2}_{d})$}, the power to obtain a
Bayes factor below a threshold $k$ can then be expressed in closed-form
\begin{align}
  \label{eq:pnlBF}
  \Pr\left(\mathrm{BF}_{01} \leq k \mid n, \mu_{d}, \tau_{d}\right) =
  \Phi\left(-\sqrt{Y} - A\right) + \Phi\left(-\sqrt{Y} + A\right)
\end{align}
with
\begin{align*}
  &Y = \left(2 \mathrm{W}_{0}\left[\frac{\{1 + (n\tau^{2})/\sigma^{2}_{\scriptscriptstyle \hat{\theta}}\}^{3/2}\sqrt{e}}{2k}\right] - 1\right) \, \left\{\frac{1 + \sigma^{2}_{\scriptscriptstyle \hat{\theta}}/(n\tau^{2})}{1 + (n \tau^{2}_{d})/\sigma^{2}_{\scriptscriptstyle \hat{\theta}}} \right\}&
&\text{and}&
  &A = \frac{\mu_{d} - \theta_{0}}{\sqrt{\tau^{2}_{d} + \sigma^{2}_{\scriptscriptstyle \hat{\theta}}/n}}&
\end{align*}
with $\mathrm{W}_{0}$ the principal branch of the Lambert W function, see
Appendix~\ref{app:nmprior} for details. Based on the power
function~\eqref{eq:pnlBF}, numerical root-finding can again be used to determine
the sample size such that a specified target power is achieved. Methods for
power and sample size calculations for normal moment prior Bayes factors are
implemented in our R package (functions \texttt{nmbf01} and
\texttt{powernmbf01}).

\begin{figure}[!bt]
\begin{knitrout}
\definecolor{shadecolor}{rgb}{0.969, 0.969, 0.969}\color{fgcolor}
\includegraphics[width=\maxwidth]{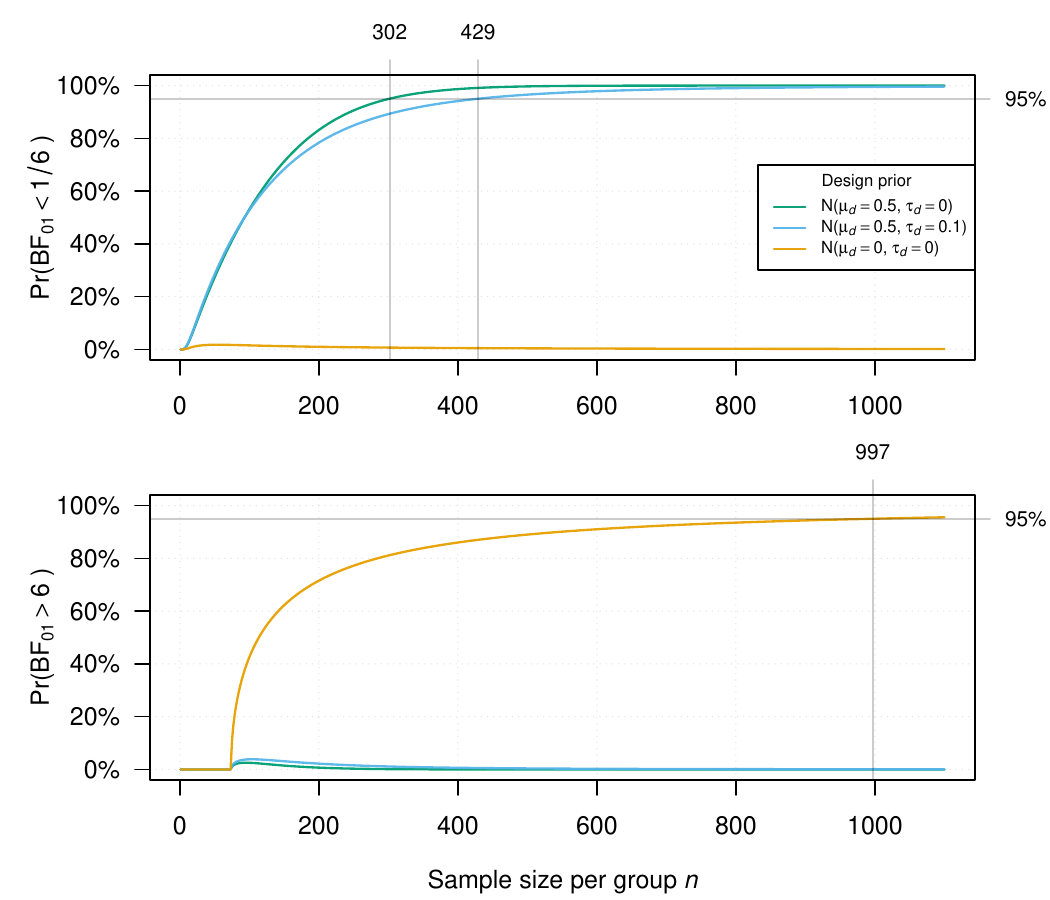} 
\end{knitrout}
\caption{Power and sample size calculations for standardized mean difference
  example from \citet[p.~133]{Schoenbrodt2017} based on normal moment prior
  Bayes factor. The normal moment analysis prior has modes at a medium effect
  size ($\tau = 0.5/\sqrt{2}$). The power is computed under
  either a point prior at a medium effect size (green), a normal prior that
  incorporates additional parameter uncertainty (blue), or under the null
  hypothesis (orange).}
\label{fig:nonlocalexample}
\end{figure}

Figure~\ref{fig:nonlocalexample} shows again power and sample size calculations
for the example from \citet{Schoenbrodt2017} considered earlier. This time,
however, we specify a normal moment analysis prior with modes at a medium
standardized mean difference effect size of $\theta = \pm 0.5$, which translates
to a prior spread parameter of
$\tau = 0.5/\sqrt{2} = 0.35$. Using the same
design priors as before, we can see that the sample sizes required to achieve a
target power of 95\% are larger than when using the normal
analysis prior as in Section~\ref{sec:schoenbrodtexample}. For example, under
the point design prior a sample size of $n = 302$ per group is
required while only $n = 153$ were required with the normal analysis prior
(see Figure~\ref{fig:comparisonBFDA}). At the same time, the sample size
required to find evidence in favor of the null hypothesis with a target power of
95\% (bottom plot) is drastically reduced
($n = 997$) compared to the normal analysis prior
($n = 6691$). This is expected since the evidence for a true null
hypothesis accumulates faster with normal moment priors \citep{Johnson2010}.

\section{Discussion}
\label{sec:discussion}
We presented methods for performing power and sample size calculations for the
situation when data are analyzed with Bayes factor hypothesis tests. These
methods rely on the approximate normality of parameter estimates (but not of the
underlying data), which is a common assumption underlying many methods for power
and sample size calculation. We have synthesized and extended previous
theoretical results on power functions for Bayes factors and implemented them in
an R package \texttt{bfpwr}. We also derived novel sample size formulas that are
easy-to-use, help fostering intuition, and enable understanding of theoretical
properties such as asymptotic power or (non-)existence of sample size for a
target power and design prior. Compared to commonly used simulation-based
methods, our methods are less general. However, in the setting where they are
applicable -- which includes many common scenarios, such as testing mean
differences -- they are faster, deterministic, and require no simulation
parameters to be specified. Therefore, we believe that the availability of such
methods addresses an important practical need and can help researchers design
efficient studies with minimal effort.

An important issue in any Bayesian design and analysis is the choice of prior
distributions. While there are differences between point and normal priors in
terms of the closed-form availability of sample sizes (see
Table~\ref{tab:summary}), we believe that the choice between them should be
guided only by prior knowledge and the questions researchers wish to answer.
Point analysis priors seem more appropriate when researchers have precise
hypotheses to test, they can also be motivated from a likelihood perspective,
and they are easier to communicate to a non-statistical audience. In contrast,
normal analysis priors can incorporate uncertainty and may therefore be more
appropriate for exploratory settings where little is known about the phenomenon
under investigation \citep{Held2018}. Similar considerations apply to the choice
of the design prior. For example, a point design prior may be specified at a
minimally relevant parameter value if such a value can be formulated. On the
other hand, if such a value is difficult to formulate but data from a previous
study are available, a normal design prior that takes the previous estimate and
its standard error as the mean and standard deviation may be a reasonable way to
incorporate prior knowledge and uncertainty. 

A clear limitation of our methodology is the asymptotic normality assumption.
This assumption may be inappropriate for certain data or parameter types, and
may lead to an underappreciation of uncertainty and consequently an
underestimation of sample size. Simulation-based methods do not have this
shortcoming, as they can be tailored to any data distribution and analysis
method. Nevertheless, simulation methods may be intimidating or too advanced for
research workers, in which case we believe it is better to do an approximate
calculation than no calculation at all. One avenue for future work might be to
extend closed-form power and sample size calculations to more specific settings
not considered here, such as, binary outcomes.

Another limitation is the types of Bayes factors that we considered for the
analysis, which is limited to univariate parameters with normal, normal moment,
$t$, or point priors under the alternative. However, after publishing an initial
version of this article on the arXiv preprint server, \citet{Wong2024} have
built on our results and extended the root-finding approach from
Section~\ref{sec:extensions} to more flexible design prior distributions and
$t$~likelihoods. Nevertheless, the approach could even be further extended to
ANOVA or regression settings.

Furthermore, in the case of two-group parameters, we have assumed equal
allocation of observations between the two groups, but unequal allocation may
sometimes be desirable. For example, it might be easier to recruit patients for
a study if they have a higher chance of receiving a new treatment than placebo,
as participation will then be more attractive. In the case of (standardized)
mean difference parameters, this is usually addressed by introducing an
allocation ratio $r = n_2/n_1$ and changing the sample size per group $n$ based
on equal allocation to $n_1 = (2n)/(1 + r)$ and $n_2 = (2n)/(1 + 1/r)$, see
e.g., \citet[p.~21]{Kieser2020}. While this adjustment has been developed for
frequentist sample size determination, it is equally applicable to the Bayes
factor methods considered here. However, for other parameter types, such as log
hazard ratios, this is more complicated \citep[p.74-75]{Kieser2020}.

We also did not consider sequential designs, where data are collected
continuously until compelling evidence for one of the competing hypotheses is
found \citep{Wald1947}, or the sample size is `recalculated' after an interim
analysis \citep[see part III in~][]{Kieser2020}. The sequential approach is
particularly attractive for Bayes factor inference as design and analysis prior
distributions can be updated based on the accumulating data. Researchers can
then make informed decisions about whether or not it is worthwhile to continue
collecting data or to stop \citep{Stefan2022}. For these purposes, it would be
interesting to consider the Bayes factor indexed by the sample size as a
stochastic process, and study its properties.

Finally, in situations where researchers have only a fixed sample size at their
disposal, it may be interesting to use a `reverse-Bayes' approach
\citep{Held2021b} and determine the design prior required to achieve a desired
target power. Researchers can then reason whether or not this prior is
scientifically plausible and they should undertake the study based on their
limited resources. This is similar to classical power analysis, where one can
determine the value of the true parameter needed to achieve a given power for a
given sample size, and then reason about whether or not this true parameter is
realistic. In both cases, however, it is important that the focus is on
reasoning about the plausibility of the design prior/parameter, and researchers
should be careful not to give the impression that this choice has been made
\textit{a priori}.

\section*{Acknowledgments}
We thank \anonymize{Angelika Stefan} and \anonymize{František Bartoš} for
valuable comments on drafts of the manuscript. We thank \anonymize{Tsz Keung
  Wong} for helpful suggestions related to our R package. We thank two anonymous
reviewers and an associate editor for constructive comments. The acknowledgment
of these individuals does not imply their endorsement of the paper.

\section*{Conflict of interest}
We declare no conflict of interest.

\section*{Software and data}
Code and data to reproduce our analyses are openly available at
\anonymize{\url{https://github.com/SamCH93/bfpwr}}. A snapshot of the repository
at the time of writing is available at
\anonymize{\url{https://doi.org/10.5281/zenodo.12582277}}. We used the
statistical programming language R version 4.4.1 (2024-06-14) for
analyses \citep{R} along with the \texttt{BFDA} \citep{Schoenbrodt2019},
\texttt{lamW} \citep{Adler2015}, \texttt{xtable} \citep{Dahl2019}, and
\texttt{knitr} \citep{Xie2024} packages.

\begin{appendices}

\section{The bfpwr R package}
\label{app:pkg}
Our R package can be installed by running \texttt{install.packages("bfpwr")} in
an R session, the development version can be found on GitHub
(\anonymize{\url{https://github.com/SamCH93/bfpwr}}). The workhorse function of
our R package is \texttt{powerbf01}. It is inspired by the \texttt{power.t.test}
function from the \texttt{stats} package, with which many user will be familiar.
As \texttt{power.t.test}, the function \texttt{powerbf01} assumes that the data
are continuous and that the parameter of interest is either a mean or a
(standardized) mean difference. The functions \texttt{pbf01} and \texttt{nbf01}
are more general and can be used for any approximately normally distributed
parameter estimate with approximate variance $\mathrm{Var}(\hat{\theta}) =
\sigma^{2}_{\scriptscriptstyle \hat{\theta}}/n$, although users have to specify
the unit standard deviation $\sigma_{\scriptscriptstyle \hat{\theta}}$
themselves. Similar functions exist also for $t$-test Bayes factors
(\texttt{powertbf01}) and normal moment prior Bayes factors
(\texttt{powernmbf01}). The following code chunk illustrate how
\texttt{powerbf01} can be used.

\begin{spacing}{1}
\begin{knitrout}
\definecolor{shadecolor}{rgb}{0.969, 0.969, 0.969}\color{fgcolor}\begin{kframe}
\begin{alltt}
\hlkwd{library}\hldef{(bfpwr)}

\hlcom{## BF parameters}
\hldef{k} \hlkwb{<-} \hlnum{1}\hlopt{/}\hlnum{6} \hlcom{# set BF_01 threshold to 1/6}
\hldef{null} \hlkwb{<-} \hlnum{0} \hlcom{# set null value to 0}
\hldef{sd} \hlkwb{<-} \hlnum{1} \hlcom{# set standard deviation of one observation to 1}
\hldef{pm} \hlkwb{<-} \hlnum{0} \hlcom{# set analysis prior mean to 0}
\hldef{psd} \hlkwb{<-} \hlkwd{sqrt}\hldef{(}\hlnum{2}\hldef{)} \hlcom{# set analysis prior SD set to sqrt(2)}
\hldef{type} \hlkwb{<-} \hlsng{"two.sample"} \hlcom{# set test to two-sample}

\hlcom{## design prior}
\hldef{dpm} \hlkwb{<-} \hlnum{0.5} \hlcom{# set design prior mean equal to medium SMD effect size}
\hldef{dpsd} \hlkwb{<-} \hlnum{0.1} \hlcom{# set positive design prior SD to incorporate parameter uncertainty}

\hlcom{## determine sample size to achieve 85% power}
\hldef{power} \hlkwb{<-} \hlnum{0.85} \hlcom{# target power}
\hldef{ssd} \hlkwb{<-} \hlkwd{powerbf01}\hldef{(}\hlkwc{k} \hldef{= k,} \hlkwc{power} \hldef{= power,} \hlkwc{sd} \hldef{= sd,} \hlkwc{null} \hldef{= null,} \hlkwc{pm} \hldef{= pm,} \hlkwc{psd} \hldef{= psd,}
                 \hlkwc{dpm} \hldef{= dpm,} \hlkwc{dpsd} \hldef{= dpsd,} \hlkwc{type} \hldef{= type)}
\hldef{ssd}
\end{alltt}
\begin{verbatim}
## 
##      Two-sample z-test Bayes factor power calculation 
## 
##                         n = 148.5498
##                     power = 0.85
##                        sd = 1
##                      null = 0
##       analysis prior mean = 0
##         analysis prior sd = 1.414214
##         design prior mean = 0.5
##           design prior sd = 0.1
##            BF threshold k = 1/6
## 
## NOTE: BF oriented in favor of H0 (BF < 1 indicates evidence for H1 over H0)
##       n is number of *observations per group*
##       sd is standard deviation of one observation (assumed equal in both groups)
\end{verbatim}
\begin{alltt}
\hlcom{## plot power curve}
\hlkwd{plot}\hldef{(ssd,} \hlkwc{nlim} \hldef{=} \hlkwd{c}\hldef{(}\hlnum{1}\hldef{,} \hlnum{400}\hldef{))}
\end{alltt}
\end{kframe}
\includegraphics[width=\maxwidth]{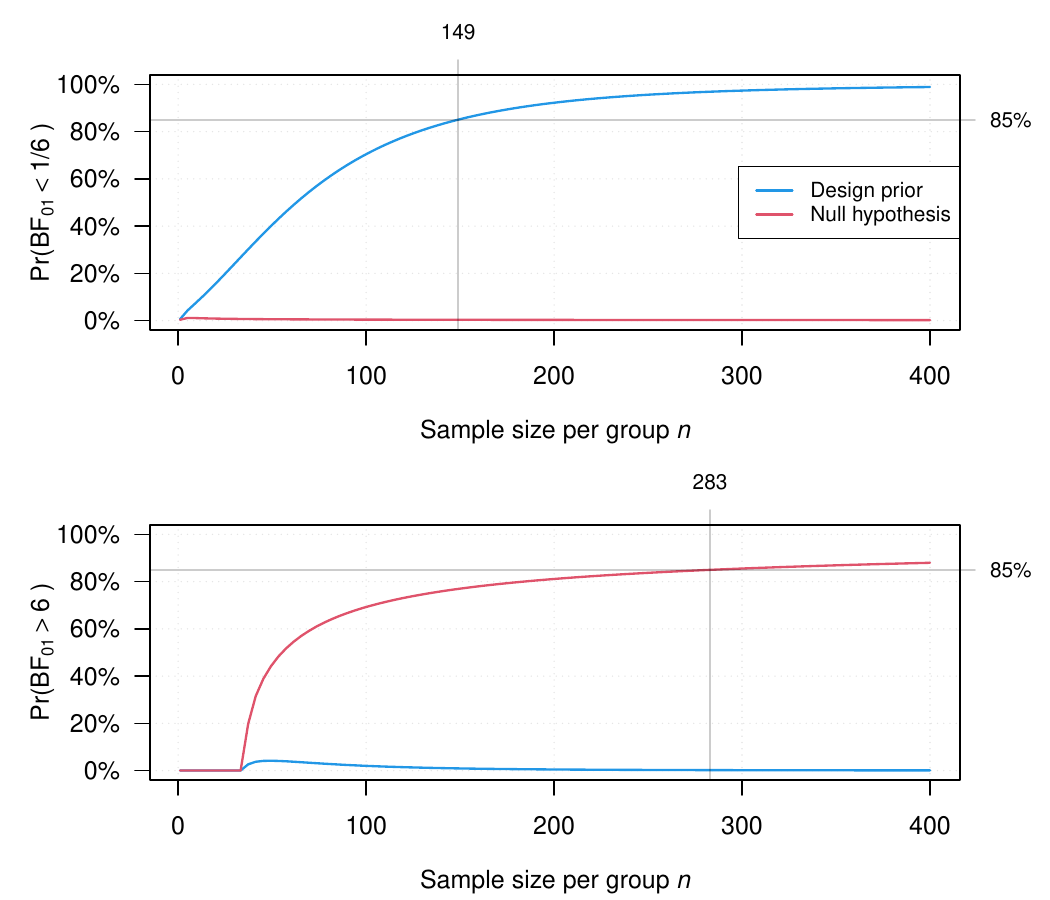} 
\end{knitrout}
\end{spacing}

\section{Distribution of the Bayes factor}
\label{app:distributions}
The Bayes factor~\eqref{eq:BF01} with point analysis prior ($\tau = 0$) can be
rewritten as
\begin{align}
\label{eq:LRdesignform}
    \text{BF}_{01} 
    &= \exp\left[\frac{n}{\sigma^2_{\scriptscriptstyle \hat{\theta}}} \left\{\hat{\theta}(\theta_0 - \mu) - \frac{\theta_0^2 - \mu^2}{2}\right\}\right].
\end{align}
Suppose that compelling evidence for $H_1$ is achieved when
$\text{BF}_{01} \leq k$. In this case, $\text{BF}_{01} \leq k$ can be rewritten
as
\begin{align*}
    \hat{\theta}(\theta_0 - \mu) \leq \frac{\sigma^2_{\scriptscriptstyle \hat{\theta}}\log k}{n} + \frac{\theta_0^2 - \mu^2}{2}.
\end{align*}
Dividing by $(\theta_0 - \mu)$ changes the inequality if $\mu > \theta_0$. We
then have that under a normal distribution
$\hat{\theta} \mid n, \mu_{d}, \tau_{d} \sim \mathrm{N}(\mu_{d}, \tau^{2}_{d} + \sigma^2_{\scriptscriptstyle \hat{\theta}}/n)$,
the probability of compelling evidence is given by~\eqref{eq:prLR}.

The Bayes factor~\eqref{eq:BF01} with normal analysis prior ($\tau > 0$) can
be rewritten as
\begin{align}
\label{eq:BFdesignform}
    \text{BF}_{01} 
    &= \sqrt{1 + \frac{n \tau^2}{\sigma^2_{\scriptscriptstyle \hat{\theta}}}} \, \exp\left(-\frac{1}{2} \left[\frac{\{\hat{\theta} - \theta_0 - \frac{\sigma^2_{\scriptscriptstyle \hat{\theta}}}{n\tau^2}(\theta_0 - \mu)\}^2}{\frac{\sigma^2_{\scriptscriptstyle \hat{\theta}}}{n}(1 + \frac{\sigma^2_{\scriptscriptstyle \hat{\theta}}}{n\tau^2})} - \frac{(\theta_0 - \mu)^2}{\tau^2}\right]\right).
\end{align}
Suppose that compelling evidence for $H_1$ is achieved when
$\text{BF}_{01} \leq k$, which can be rearranged to
\begin{align*}
    \left\{\hat{\theta} - \theta_0 - \frac{\sigma^2_{\scriptscriptstyle \hat{\theta}}}{n\tau^2}(\theta_0 - \mu)\right\}^2 \geq
    \left\{\log\left(1 + \frac{n\tau^2}{\sigma^2_{\scriptscriptstyle \hat{\theta}}}\right) + \frac{(\theta_0 - \mu)^2}{\tau^2} - \log k^2\right\} \left(1 + \frac{\sigma^2_{\scriptscriptstyle \hat{\theta}}}{n\tau^2} \right) \frac{\sigma^2_{\scriptscriptstyle \hat{\theta}}}{n}.
\end{align*}
Therefore, under a normal distribution
$\hat{\theta} \mid n, \mu_{d}, \tau_{d} \sim \mathrm{N}(\mu_{d}, \tau^{2}_{d} + \sigma^2_{\scriptscriptstyle \hat{\theta}}/n)$,
the probability of compelling evidence is given by~\eqref{eq:prBF}.

\section{Limiting power of Bayes factor with normal analysis prior}
\label{app:asymptotics}
We have that
\begin{align*}
  \lim_{n\to \infty} M
  = \frac{\mu_{d} - \theta_{0}}{\tau_{d}}
\end{align*}
and
\begin{align*}
  \lim_{n\to \infty} X
  = \lim_{n \to \infty} \left[\left\{\log\left(1 + \frac{n\tau^2}{\sigma^2_{\scriptscriptstyle \hat{\theta}}}\right) + \frac{(\theta_0 - \mu)^2}{\tau^2} - \log k^2\right\}
  \frac{\sigma^2_{\scriptscriptstyle \hat{\theta}}}{n\tau^{2}_{d} + \sigma^{2}_{\scriptscriptstyle \hat{\theta}}}\right].
\end{align*}
Thus, when also $\tau_{d} \downarrow 0$ and $\mu_{d} \neq \theta_{0}$, both $M$
and $X$ diverge but the $M$ term diverges faster than the $X$ term. When
$\tau_{d} > 0$, the $M$ term approaches a constant while the $X$ term approaches
zero. Consequently, in both cases it holds that
\begin{align*}
  \lim_{n\to \infty}  \Pr(\text{BF}_{01} \leq k \mid n, \mu_{d} , \tau_{d}, \tau > 0)
  &= \lim_{n\to \infty} \left\{\Phi(-\sqrt{X} - M) + \Phi(-\sqrt{X} + M)\right\} 
  = 1.
\end{align*}

\section{Sample size for Bayes factor with local normal prior}
\label{app:lambertWderiv}

Equating the power function~\eqref{eq:prBFcenter} to $1 - \beta$ and applying
algebraic manipulations, we have that
\begin{align*}
  z^{2}_{(1 - \beta)/2}
  &= \left\{\log\left(1 + \frac{n\tau^2}{\sigma^2_{\scriptscriptstyle \hat{\theta}}}\right) - \log k^2\right\} \frac{\sigma^2_{\scriptscriptstyle \hat{\theta}}}{n\tau^2} \\
  &\approx \left\{\log\left(\frac{n\tau^2}{\sigma^2_{\scriptscriptstyle \hat{\theta}}}\right) - \log k^2\right\} \frac{\sigma^2_{\scriptscriptstyle \hat{\theta}}}{n\tau^2} \\
  &= \log\left(\frac{n\tau^2}{\sigma^2_{\scriptscriptstyle \hat{\theta}}k^{2}}\right)  \frac{\sigma^2_{\scriptscriptstyle \hat{\theta}}}{n\tau^2}
\end{align*}
Multiplying by $-k^{2}$ and rewriting the second factor on the right-hand-side
as exponential leads to
\begin{align*}
 -k^{2} \, z^{2}_{(1 - \beta)/2}
  &= -\log\left(\frac{n\tau^2}{\sigma^2_{\scriptscriptstyle \hat{\theta}}k^{2}}\right)   \exp\left\{-\log\left(\frac{n\tau^2}{\sigma^2_{\scriptscriptstyle \hat{\theta}}k^{2}}\right) \right\}.
\end{align*}
Hence, we can apply the Lambert W function to obtain
\begin{align*}
  -\log\left(\frac{n\tau^2}{\sigma^2_{\scriptscriptstyle \hat{\theta}}k^{2}}\right)   = \mathrm{W}\left(-k^{2} \, z^{2}_{(1 - \beta)/2}\right)
\end{align*}
from which we obtain the sample size
\begin{align*}
 n = \frac{\sigma^2_{\scriptscriptstyle \hat{\theta}}}{\tau^{2}} \, k^{2} \, \exp \left\{ -\mathrm{W}\left(-k^{2} \, z^{2}_{(1 - \beta)/2}\right)\right\}.
\end{align*}
For arguments $y \in (-1/e, 0)$ , the Lambert W function has two branches. The
sample size is obtained from the branch commonly denoted as
$\mathrm{W}_{-1}(\cdot)$ which satisfies $\mathrm{W}(x) < -1$ for
$y \in (-1/e, 0)$ \citep{Corless1996}. This is because this branch always leads
to larger sample sizes than the other and guarantees that unit information
sample sizes are always larger than one.

\section{Simulation-based evaluation of the bfpwr package}
\label{app:simeval}
\begin{figure}[!htb]

\begin{knitrout}
\definecolor{shadecolor}{rgb}{0.969, 0.969, 0.969}\color{fgcolor}
\includegraphics[width=\maxwidth]{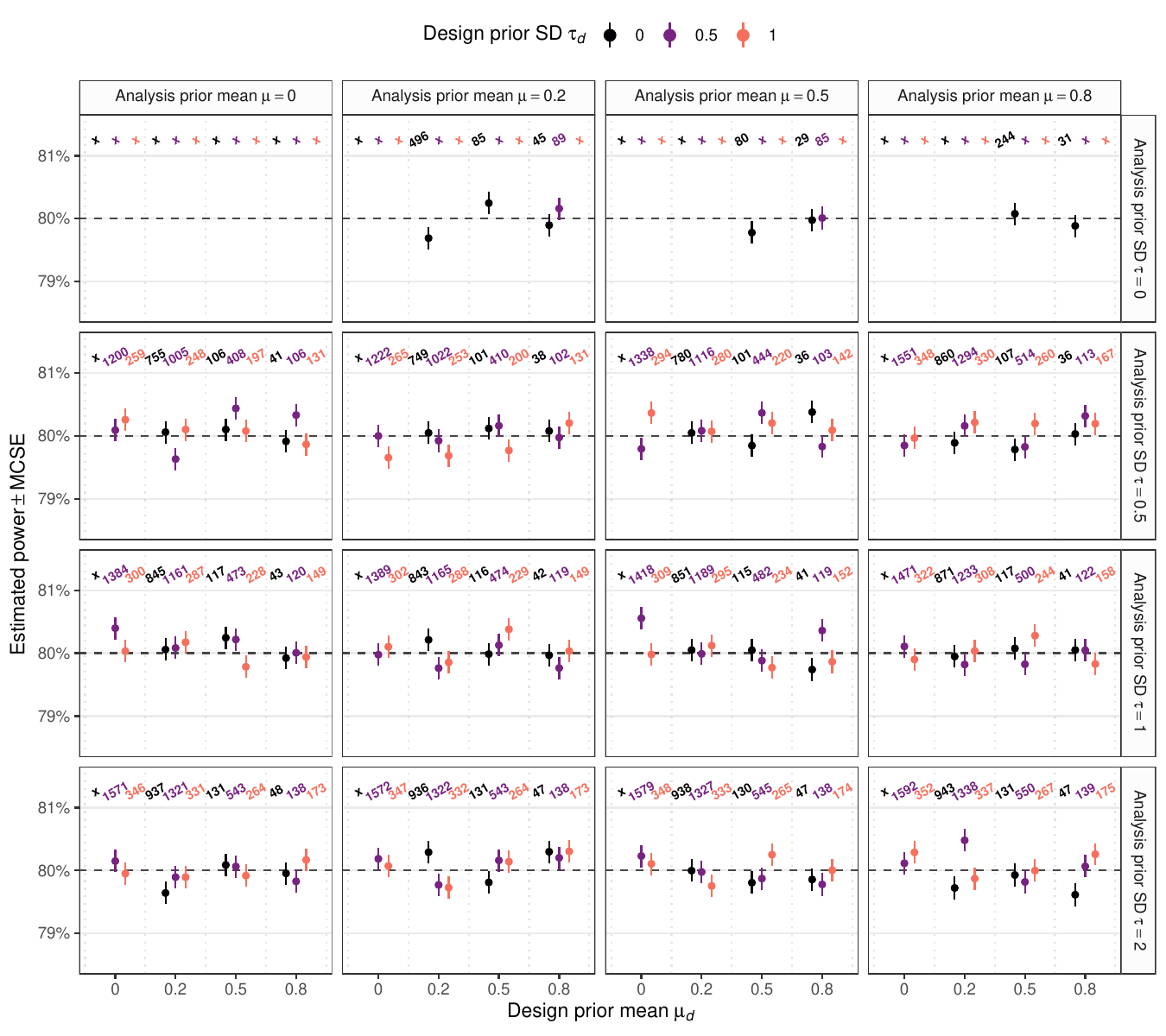} 
\end{knitrout}
\caption{Simulation-based evaluation of power and sample size calculations
  related to the Bayesian $z$-test as implemented in the \texttt{bfpwr} package.
  The top of each plot shows the sample size to obtain a Bayes factor equal or
  below $k = 1/10$ with a target power of 80\%
  for the corresponding combination of analysis and design prior (an ``x'' is
  shown if the target power is impossible to achieve for a given condition).
  50'000 Bayes factors were then simulated based
  on this sample size, from which then the power was empirically estimated.}
\label{fig:simeval}
\end{figure}

Figure~\ref{fig:simeval} shows a simulation-based evaluation of the power and
sample size calculation methods for the Bayesian $z$-test as implemented in our
\texttt{bfpwr} R package. The values for the design and analysis prior means
were chosen to represent conventions for no (0), small (0.2), medium (0.5), and
large (0.8) standardized mean differences \citep{Cohen1992}. The null and
alternative hypotheses were defined as $H_{0}\colon \theta = 0$ against $H_{1}
\colon \theta \neq 0$. The standard deviations were chosen to include point and
normal priors. For each combination of analysis/design prior mean/standard
deviation, the sample size to obtain a Bayes factor equal or below $k =
1/10$ with a target power of 80\% was computed
(shown at the top of each plot). This sample size along with the design prior
was subsequently used to simulate 50'000
standardized mean difference parameter estimates based on which
50'000 Bayes factors were computed. The power was
then estimated from the proportion of Bayes factors equal or below the level $k
= 1/10$. Note that for certain design/analysis prior combinations, it
is impossible to achieve the target power with a finite sample size. In this
case an ``x'' is shown in the plot.

We can see that in all conditions, the simulation-based estimate of the power
closely spreads around the target power of 80\%. The
maximally observed discrepancy is 0.56\% while the median discrepancy is
0.14\%. This
suggests that the power and sample size calculation methods work as intended.

\section{Power with normal moment prior}
\label{app:nmprior}

Setting the Bayes factor~\eqref{eq:nlBF} to less or equal than $k$ and applying
algebraic manipulations, we can bring the inequality into the form
\begin{align*}
  \exp\left[1 + \frac{n(\hat{\theta} - \theta_{0})^{2}}{
  \sigma^{2}_{\scriptscriptstyle \hat{\theta}}\{1 + \sigma^{2}_{\scriptscriptstyle \hat{\theta}}/(n \tau^{2})\}}\right] \,
  \left[1 + \frac{n(\hat{\theta} - \theta_{0})^{2}}{
  \sigma^{2}_{\scriptscriptstyle \hat{\theta}}\{1 + \sigma^{2}_{\scriptscriptstyle \hat{\theta}}/(n \tau^{2})\}}\right] \geq
  \frac{\{1 + (n\tau^{2})/\sigma^{2}_{\scriptscriptstyle \hat{\theta}}\}\sqrt{e}}{2k}.
\end{align*}
Applying the Lambert W function on both sides, leads to
\begin{align}
  \label{eq:powernmp2}
  1 + \frac{n(\hat{\theta} - \theta_{0})^{2}}{\sigma^{2}_{\scriptscriptstyle \hat{\theta}}\{1 + \sigma^{2}_{\scriptscriptstyle \hat{\theta}}/(n \tau^{2})\}} \geq
  \mathrm{W}_{0} \left[\frac{\{1 + (n\tau^{2})/\sigma^{2}_{\scriptscriptstyle \hat{\theta}}\}\sqrt{e}}{2k}\right].
\end{align}
Since the argument of the Lambert W function is real and always non-negative,
only the principal branch $\mathrm{W}_{0}$ can satisfy the inequality. Assuming a
\mbox{$\hat{\theta} \mid n, \mu_{d}, \tau_{d} \sim \mathrm{N}(\mu_{d}, \tau^{2}_{d} + \sigma^2_{\scriptscriptstyle \hat{\theta}}/n)$}
distribution induced by a normal design prior, we can rearrange the
inequality~\eqref{eq:powernmp2} and obtain the power function~\eqref{eq:pnlBF}.

\end{appendices}

{\small
\bibliographystyle{apalikedoiurl}
\bibliography{bibliography}
}

\end{document}